\documentclass[superscriptaddress,groupedaddress,nofootnoteinbib,12pt]{article}
\usepackage{color} \usepackage{graphicx} \usepackage{dcolumn}
\usepackage{bm} \usepackage{bbm} \usepackage{amssymb}
\usepackage{amsmath} \usepackage{sectsty} \usepackage{colortbl}
\usepackage{latexsym} \usepackage{float} \usepackage{ifthen}
\usepackage{caption,subfig} \usepackage{enumerate} \usepackage{url}
\usepackage{caption,subfig} \usepackage{jcappub}

\newcommand{\calO}{\mathcal{O}}

\newcommand{\eff}{{\rm eff}}

\newcommand{\Id}{\mathbbm 1}

\def\be{\begin{equation}}
\def\ee{\end{equation}}
\def\ba{\begin{eqnarray}}
\def\ea{\end{eqnarray}}
\def\beq{\begin{eqnarray}}
\def\eeq{\end{eqnarray}}

\def\L*{{\cal L}_*}
\def\L{\mathcal{L}}
\def\({\left(}
\def\){\right)}

\def\<{\langle}
\def\>{\rangle}

\def\cs2{c_{s}^{2}}

\def\be{\begin{equation}}
\def\ee{\end{equation}}
\def\ba{\begin{eqnarray}}
\def\ea{\end{eqnarray}}
\def\beq{\begin{eqnarray}}
\def\eeq{\end{eqnarray}}

\def\L*{{\cal L}_*}
\def\L{\mathcal{L}}
\def\({\left(}
\def\){\right)}

\def\<{\langle}
\def\>{\rangle}

 \def\be   {\begin{equation}}   \def\ee   {\end{equation}}

 \def\ba  {\begin{eqnarray}}   \def\ea  {\end{eqnarray}}

\renewcommand{\thesubsection}{\arabic{section}.\arabic{subsection}}

\usepackage{ulem}
\newcommand{\ud}{\mathrm{d}} 

\begin{document}
\hspace{5.2in} \mbox{NORDITA-2015-57}\\\vspace{-1.03cm} 

\title{Dipolar Dark Matter \\with Massive Bigravity}

\author{Luc Blanchet$^{a}$ and Lavinia Heisenberg$^{b,c}$}
\affiliation{$^{a}$
  $\mathcal{G}\mathbb{R}\varepsilon{\mathbb{C}}\mathcal{O}$ Institut
  d'Astrophysique de Paris --- UMR 7095 du CNRS, Universit\'e Pierre
  \& Marie Curie, 98\textsuperscript{bis} boulevard Arago, 75014
  Paris, France} \affiliation{$^{b}$ Nordita, KTH Royal Institute of
  Technology and Stockholm University, Roslagstullsbacken 23, 10691
  Stockholm, Sweden} \affiliation{$^{c}$ Department of Physics \& The
  Oskar Klein Centre, AlbaNova University Centre, 10691 Stockholm,
  Sweden}

\emailAdd{blanchet@iap.fr} 
\emailAdd{laviniah@kth.se}

\abstract{Massive gravity theories have been developed as viable IR
  modifications of gravity motivated by dark energy and the problem of
  the cosmological constant. On the other hand, modified gravity and
  modified dark matter theories were developed with the aim of solving
  the problems of standard cold dark matter at galactic scales. Here
  we propose to adapt the framework of ghost-free massive bigravity
  theories to reformulate the problem of dark matter at galactic
  scales. We investigate a promising alternative to dark matter called
  dipolar dark matter (DDM) in which two different species of dark
  matter are separately coupled to the two metrics of bigravity and
  are linked together by an internal vector field. We show that this
  model successfully reproduces the phenomenology of dark matter at
  galactic scales (\textit{i.e.} MOND) as a result of a mechanism of
  gravitational polarisation. The model is safe in the gravitational
  sector, but because of the particular couplings of the matter
    fields and vector field to the metrics, a ghost in the decoupling
    limit is present in the dark matter sector. However, it might be possible
    to push the mass of the ghost beyond the strong coupling scale by
    an appropriate choice of the parameters of the model. Crucial questions to
    address in future work are the exact mass of the ghost, and the
    cosmological implications of the model.}

\maketitle

\section{Introduction}

In the last century, cosmology has progressively developed from a
philosophical to an empirical scientific discipline, witnessing high
precision cosmological observations, which culminated with the
standard model of cosmology, the $\Lambda$-CDM model~\cite{OS95}. The
standard model is based on General Relativity (GR) and is in great
agreement with the abundance of the light elements in the big bang
nucleosynthesis, the anisotropies of the cosmic microwave background
(CMB), the baryon acoustic oscillations, lensing and the observed
large scale structures. It notably relies on the presence of
dark energy in form of a cosmological constant $\Lambda$, giving rise
to the accelerated expansion of the universe.

\subsection{Massive gravity context}

In the prevailing view, the cosmological constant corresponds to the
constant vacuum energy density which receives large quantum
corrections. Unfortunately, so far there is no successful mechanism to
explain the observed unnatural tiny value of the cosmological
constant, see \textit{e.g.}~\cite{Padmanabhan,Martin}. This
difficulty has sparked a whole industry studying modifications of
gravity in the infra-red (IR) invoking new dynamical degrees of
freedom.

One tempting route is massive gravity, motivated by the possibility
that the graviton has a mass.  It was a challenge over forty years to
construct a covariant non-linear theory for massive gravity. The
foregathered remarkable amount of effort finally rose to the
challenge~\cite{deRham:2010ik,deRham:2010kj,deRham11,Hassan:2011hr,Hassan:2011vm,Hassan:2011tf,Hassan:2011ea,Hassan:2012qv},
to construct the potential interactions in a way that got rid of the
Boulware--Deser (BD) ghost~\cite{Boulware:1973my}. This theory has
been further extended to more general models by adding additional
degrees of freedom. A very important extension is massive bigravity
theory~\cite{Hassan:2011zd}.

Once the interactions in the gravitational sector were guaranteed to
be ghost-free, a natural follow up question was how to couple this
theory to the matter fields without spoiling the
ghost-freedom. First attempts were already discussed in the
original paper of bigravity~\cite{Hassan:2011zd}. If one couples the
matter fields to both metrics simultaneously, this reintroduces the BD
ghost~\cite{Yamashita:2014fga,deRham:2014naa}. Furthermore, the
one-loop quantum corrections detune the special potential interactions
at an unacceptable low scale and hence this way of coupling is not a
consistent one~\cite{deRham:2014naa}. The safer way is to couple the
matter fields to just one metric. In this way, the quantum corrections
give rise to contributions in form of cosmological constants. One
could also try to couple the matter field to the massless mode, which
is unfortunately also not ghost free~\cite{Hassan:2012wr}. Another
possible way of coupling the matter fields to both metrics is through
a composite effective metric constructed out of both
metrics~\cite{deRham:2014naa,deRham:2014fha,Noller:2014sta,LH15},
which is unique in the sense that it is the only non-minimal matter
coupling that maintains ghost-freedom in the decoupling
limit~\cite{deRham:2015cha, Huang:2015yga, Heisenberg:2015iqa}.
Furthermore, the quantum corrections are guaranteed to maintain the
nice potential structure. Other important consequence of this new
matter coupling is the fact that it helps to evade the no-go
result~\cite{PhysRevD.84.124046} for the flat
Friedmann--Lema\^itre--Robertson--Walker (FLRW) background. A detailed
perturbed ADM analysis revealed the existence of a BD ghost
originating from an operator involving spatial
derivatives~\cite{deRham:2014naa,deRham:2014fha}. Therefore, the ghost
will probably reappear for highly anisotropic solutions. Since the
ghost remains absent up to the strong coupling scale, the matter
coupling can be considered in an effective field theory sense at
the very least till the strong coupling scale. The precise cut-off of
the theory or mass of the ghost has still to be established.

The absence of the BD ghost is not only important at the classical
level, but also at the quantum level. For the classical ghost-freedom
the relative tuning of the potential interactions is the key
point. Therefore one has to ensure that the quantum corrections do not
detune the potential interactions. Concerning the decoupling limit, it
is easy to show that the theory receives no quantum corrections
\textit{via} the non-renormalization theorem due to the antisymmetric
structure of the interactions~\cite{deRham:2012ew} (in fact the same
antisymmetric structure of the Galileon interactions protect them from
quantum
corrections~\cite{Luty:2003vm,Nicolis:2004qq,Heisenberg:2014raa}). Beyond
the decoupling limit, the quantum corrections of the matter loops
maintain the potential interactions provided that the above criteria
are fulfilled. Concerning the graviton loops, they do destroy the
relative tuning of the potential interactions. Nevertheless, this
detuning is harmless since the mass of the corresponding BD ghost is
never below the cut-off scale of the theory~\cite{deRham:2013qqa}. The
bimetric version of the theory shares the same
property~\cite{Heisenberg:2014rka}.

Massive gravity/bigravity theory has a rich phenomenology.  The
decoupling limit admits stable self-accelerating
solutions~\cite{deRham:2010tw}, where the helicity-0 degree of freedom
of the massive graviton plays the role of a condensate whose energy
density sources self-acceleration. Unfortunately these solutions
suffer from strong coupling issues due to the vanishing kinetic term
of the vector modes~\cite{deRham:2010tw,Tasinato:2012ze}. With the
original massive gravity and the restriction of flat reference metric,
one faces the no-go result, namely that there are no flat FLRW
solution~\cite{PhysRevD.84.124046}. One can construct
self-accelerating open FLRW solutions~\cite{Gumrukcuoglu:2011ew},
which however have three instantaneous
modes~\cite{Gumrukcuoglu:2011zh} and suffer from a nonlinear ghost
instability~\cite{PhysRevLett.109.171101}. The attempt to promote the
reference metric to de Sitter~\cite{Fasiello:2012rw,Langlois:2012hk}
also failed due to the presence of the Higuchi
ghost~\cite{Fasiello:2012rw}. In order to avoid these difficulties,
one either gives up on the FLRW symmetries~\cite{PhysRevD.84.124046},
or invokes new additional degrees of
freedom~\cite{Huang:2012pe,Hassan:2011zd,Mukohyama:2014rca}.

Thanks to the freedom gained in the inclusion of the second kinetic
term in the bimetric extension~\cite{Hassan:2011zd}, there now exists
many elaborate works concerning the cosmology of the bigravity theory,
see
\textit{e.g.}~\cite{Volkov:2011an,vonStrauss:2011mq,Comelli:2014bqa}. In
the case of minimally coupled matter fields and small graviton mass,
the theory admits several interesting branches of
solutions. Unfortunately, among them the self-accelerating branch is
unstable due to the presence of three instantaneous modes, and a
second branch of solutions admits an early time gradient
instability~\cite{Comelli:2012db}. Nevertheless, there exists attempts to
overcome the gradient instability either by viable
though finely tuned solutions in the case of a strongly interacting
bimetric theory with $m\gg H_0$~\cite{DeFelice:2013nba,DeFelice:2014nja},
or by demanding that $M_g\gg M_f$ as was proposed in \cite{Akrami:2015qga}.
See also
other recent works concerning the phenomenology of bimetric
gravity~\cite{Enander:2015vja,Enander:2015kda,Cusin:2015pya,Konnig:2015lfa,
Fasiello:2015csa}.

\subsection{Dark matter context}

On a quite different but equally fascinating topic is the
phenomenology of dark matter at galactic scales. The evidence for dark
matter is through the measurement of the rotation curves of spiral
galaxies which turn out to be approximately flat, contrary to the
Newtonian prediction based on ordinary baryonic
matter~\cite{Bosma,Rubin}. The standard explanation is that the disk
of galaxies is embedded into the quasi-spherical potential of a huge
halo of dark matter, and that this dark matter is the same as the cold
dark matter (CDM) which is evidenced at large cosmological scales
notably with the fit of $\Lambda$-CDM to the CMB
anisotropies~\cite{HuD02,BHS05}. Unfortunately this explanation faces
severe challenges when compared to observations at galactic
scales~\cite{SandMcG02,FamMcG12}. There are predictions of the
$\Lambda$-CDM model that are not observed, like the phase-space
correlation of galaxy satellites and the generic formation of dark
matter cusps in the central regions of galaxies. Even worse, there are
observations which are not predicted by $\Lambda$-CDM, such as the
tight correlation between the mass discrepancy (luminous \textit{vs.}
dynamical mass) which measures the presence of dark matter and the
involved scale of acceleration, the famous baryonic Tully-Fisher (BTF)
relation for spiral galaxies~\cite{TF77,McG00,McG11}, and its
equivalent for elliptical galaxies, the Faber-Jackson
relation~\cite{Sand10}.

Instead of additional, non-visible mass, Milgrom~\cite{Milg1, Milg2,
  Milg3} proposed an amendment to the Newtonian laws of motion in
order to account for the phenomenology of dark matter in galaxies,
dubbed MOND for MOdified Newtonian Dynamics. According to the MOND
hypothesis the change has a relevant influence on the motion only for
very small accelerations, as they occur at astronomical scales, below
the critical value $a_0\simeq 1.2\times 10^{-10}\,\text{m}/\text{s}^2$
measured for instance from the BTF relation.\footnote{A striking
  observation is the numerical coincidence that
  $a_0\sim\sqrt{\Lambda}$~\cite{SandMcG02,FamMcG12}.} A more elaborate
version of MOND is the Bekenstein-Milgrom~\cite{BekM84} modification
of the Poisson equation of Newtonian gravity. All the challenges met
by $\Lambda$-CDM at galactic scales are then solved --- sometimes with
incredible success --- by the MOND
formula~\cite{SandMcG02,FamMcG12}. Unfortunately, MOND faces
difficulties in explaining the dark matter distribution at the larger
scales of galaxy clusters~\cite{GD92,PSilk05,Clowe06,Ang08,Ang09}. It
is also severely constrained by observations in the solar
system~\cite{Milg09,BN11}.

Reconciling $\Lambda$-CDM at cosmological scales and MOND at galactic
scales into a single relativistic theory is a great challenge. There
has been extensive works on this. One approach is to modify gravity by
invoking new gravitational degrees of freedom without the presence of
dark
matter~\cite{Sand97,Bek04,Sand05,ZFS07,Halle08,bimond1,BDgef11,DEW11,BM11,Arraut2014}. Primary
examples are the tensor-vector-scalar (TeVeS)
theory~\cite{Bek04,Sand05} and generalized Einstein-{\AE}ther
theories~\cite{ZFS07,Halle08}.  Another approach is a new form of dark
matter \textit{\`a la} MOND, called dipolar dark matter (DDM). It is
based on the dielectric analogy of MOND~\cite{B07mond} --- a
remarkable property of MOND with possible far-reaching
implications. Indeed the MOND equation represents exactly the
gravitational analogue (in the non-relativistic limit) of the Gauss
equation of electrostatics modified by polarisation effects in
non-linear dielectric media. Some early realizations of this analogy
were proposed in~\cite{BL08,BL09,BLLM13} and shown to also reproduce
the cosmological model $\Lambda$-CDM. The best way to interpret this
property is by a mechanism of gravitational polarisation, involving
two different species of particles coupled to different gravitational
potentials. This was the motivation for pushing forward a more
sophisticated model in the context of a bimetric extension of
GR~\cite{BBwag,BB14}. In this model the two species of dark matter
particles that couple to the two metrics are linked through an
internal vector field. The model~\cite{BB14} is able to recover MOND
at galactic scales and the cosmological model $\Lambda$-CDM at large
scales. Furthermore the post-Newtonian parameters (PPN) in the solar
system are the same as those of GR. Unfortunately, even if this model
delivers a very interesting phenomenology, it is plagued by ghost
instabilities in the gravitational sector already at the linear order
of perturbations and within the decoupling
limit~\cite{Blanchet:2015sra}. A more promising model was then
proposed in~\cite{Blanchet:2015sra}, with the hope to cure these
instabilities in the gravitational sector.

In the present work we will study in detail the theoretical and
phenomenological consistency of this recently proposed model for
dipolar dark matter in the context of bimetric
gravity~\cite{Blanchet:2015sra}. We will first work out the covariant
field equations in Sec.~\ref{sec:theoryLBLH} and analyze the linear
field equations in Sec.~\ref{sec:linear_pert}. As next, we will
investigate in Sec.~\ref{sec:polarisation_mond} the mechanism of
gravitational polarisation in the non relativistic approximation, and
show how it successfully recovers the MOND phenomenology on galactic
scales. Finally we will pay attention in Sec.~\ref{sec:decoupling} to
the matter sector and investigate the interactions in the decoupling
limit. The outcome of Sec.~\ref{sec:decoupling} is that a ghost
instability is still present in the (dark) matter sector of the
model. The paper ends with some short conclusions in
Sec.~\ref{sec:conclusion}.


\section{The covariant theory}
\label{sec:theoryLBLH} 

A new relativistic model for dipolar dark matter has been recently
proposed in~\cite{Blanchet:2015sra}. Basically this model is defined
by combining the specific dark matter sector of a previous
model~\cite{BB14} with gravity in the form of ghost-free bimetric
extensions of GR~\cite{Hassan:2011zd}. The dark matter in this model
consists of two types of particles respectively coupled to the two
dynamical metrics $g_{\mu\nu}$ and $f_{\mu\nu}$ of bigravity. In
addition, these two sectors are linked together \textit{via} an
additional internal field in the form of an Abelian $U(1)$ vector
field $\mathcal{A}_\mu$. This vector field is coupled to the effective
composite metric $g_\text{eff}$ of bigravity which is built out of the
two metrics $g$ and $f$, and which has been shown to be allowed in the
matter Lagrangian in the effective field theory sense at least up to
the strong coupling scale~\cite{deRham:2014naa,deRham:2014fha}.  The
total matter-plus-gravity Lagrangian reads\footnote{The metric
  signature convention is $(-,+,+,+)$. We adopt geometrical units with
  the gravitational constant, the Planck constant and the speed of
  light being unity, $G=\hbar=c=1$, unless specified otherwise.}
\begin{align}
\mathcal{L} &= \sqrt{-g}\biggl(\frac{M_g^2}{2}R_g
-\rho_\text{bar}-\rho_g\biggr)
+\sqrt{-f}\biggl(\frac{M_f^2}{2}R_f-\rho_f\biggr) \nonumber\\ &
+\sqrt{-g_\text{eff}} \biggl[ m^2M_\text{eff}^2+
  \mathcal{A}_\mu\bigl(j_g^\mu-j_f^\mu\bigr) + \lambda
  M_\text{eff}^2\,\mathcal{W}\bigl(\mathcal{X}\bigr)
  \biggr]\,.\label{lagrangian}
\end{align}
Here $R_g$ and $R_f$ are the Ricci scalars of the two metrics, and the
scalar energy densities of the ordinary matter modelled simply by
pressureless baryons,\footnote{We have in mind that the baryons
  actually represent the full standard model of particle physics.} and
the two species of pressureless dark matter particles are denoted by
$\rho_\text{bar}$, $\rho_g$ and $\rho_f$ respectively. In addition
$j_g^\mu$, $j_f^\mu$ stand for the mass currents of the dark matter,
defined by
%
\begin{equation}\label{currents}
j^\mu_g = \frac{\sqrt{-g}}{\sqrt{-g_\text{eff}}} \,J_g^\mu \qquad
\text{and} \qquad j^\mu_f = \frac{\sqrt{-f}}{\sqrt{-g_\text{eff}}}
\,J_f^\mu\,,
\end{equation}
where $J^\mu_g=r_g\rho_g u_g^\mu$ and $J^\mu_f=r_f\rho_f u_f^\mu$ are
the corresponding conserved dark matter currents associated with the
respective metrics $g$ and $f$,
thus satisfying $\nabla^g_\mu
J_g^\mu=0$ and $\nabla^f_\mu J_f^\mu=0$. Here $\rho_g$ and
  $\rho_f$ denote the scalar densities and $u^\mu_g$, $u^\mu_f$ are
  the four velocities normalized to $g_{\mu\nu}u_g^\mu u_g^\nu=-1$ and
  $f_{\mu\nu}u_f^\mu u_f^\nu=-1$. We also introduced two constants
  $r_g$ and $r_f$ specifying the ratios between the charge and the
  inertial mass of the dark matter particles.

The metric independent matter degrees of freedom are the
  coordinate densities $\rho^*_g=\sqrt{-g}\rho_g u^0_g$ and
  $\rho^*_f=\sqrt{-f}\rho_f u^0_f$, and the coordinate velocities
  $v^\mu_g=u^\mu_g/u^0_g$ and $v^\mu_f=u^\mu_f/u^0_f$. The associated
  metric independent currents are $J^{*\mu}_g=r_g \rho^*_g v^\mu_g$
  and $J^{*\mu}_f=r_f \rho^*_f v^\mu_f$. They are conserved in the
  ordinary sense, $\partial_\mu J^{*\mu}_g=0$ and $\partial_\mu
  J^{*\mu}_f=0$. They relate to the classical currents $J^\mu_g$,
  $J^\mu_f$ or $j^\mu_g$, $j^\mu_f$ by
$$J^{*\mu}_g = \sqrt{-g}\,J_g^\mu = \sqrt{-g_\text{eff}}\,j_g^\mu \qquad
\text{and} \qquad J^{*\mu}_f = \sqrt{-f}\,J_f^\mu =
\sqrt{-g_\text{eff}}\,j_f^\mu\,.$$
Notice that the coupling term $\sqrt{-g_\text{eff}}\,\mathcal{A}_\mu
(j_g^\mu-j_f^\mu)$ in the action~\eqref{lagrangian} is actually
independent of any metric, neither $g$ nor $f$ nor $g_\text{eff}$. We
shall study in detail in Sec.~\ref{sec:decoupling} the implications of
the term $\sqrt{-g_\text{eff}}\,\mathcal{A}_\mu (j_g^\mu-j_f^\mu)$ for
the decoupling limit of the theory.

The vector field $\mathcal{A}_\mu$ is generated by the dark matter
currents and plays the role of a ``graviphoton''. The presence of this
internal field is necessary to stabilize the dipolar medium and will
yield the wanted mechanism of gravitational polarisation, as we shall
see in Sec.~\ref{sec:polarisation_mond}. It has a non-canonical
kinetic term $\mathcal{W}(\mathcal{X})$, where $\mathcal{W}$ is a
function which is left unspecified at this stage, and
\begin{equation}\label{X}
\mathcal{X} = 
- \frac{\mathcal{F}^{\mu\nu}\mathcal{F}_{\mu\nu}}{4\lambda}\,.
\end{equation}
Here $\lambda$ is a constant and the field strength is constructed
with the effective composite metric $g^\text{eff}_{\mu\nu}$ given
by~\eqref{effmetric} below, \textit{i.e.} $\mathcal{F}^{\mu\nu} =
g_\text{eff}^{\mu\rho}
g_\text{eff}^{\nu\sigma}\mathcal{F}_{\rho\sigma}$ with
$\mathcal{F}_{\mu\nu} = \partial_\mu\mathcal{A}_\nu -
\partial_\nu\mathcal{A}_\mu$ not depending on the metric as usual.

The model~\eqref{lagrangian} is defined by several constant
parameters, the coupling constants $M_g$, $M_f$ and $M_\text{\eff}$,
the mass of the graviton $m$, the charge to mass ratios $r_g$ and
$r_f$ of dark matter, the constant $\lambda$ associated with the
vector field, and the arbitrary constants $\alpha$ and $\beta$
entering the effective metric~\eqref{effmetric}. Parts of these
constants, as well as the precise form of the function
$\mathcal{W}(\mathcal{X})$, will be determined in
Sec.~\ref{sec:polarisation_mond} when we demand that the model
reproduces the required physics of dark matter at galactic scales. In
particular $\lambda$ will be related to the MOND acceleration scale
$a_0$, and the model will finally depend on $a_0$ and the mass
of the graviton $m$. Additionally, there will be still some remaining
freedom in the parameters $\alpha$ and $\beta$, and in the
  coupling constants $M_g$ and $M_f$.

In the model~\eqref{lagrangian} the ghost-free potential interactions
between the two metrics $g$ and $f$ take the particular form of the
square root of the determinant of the effective
metric~\cite{deRham:2014naa,deRham:2014fha,LH15}
\begin{equation}\label{effmetric}
g^\text{eff}_{\mu\nu}=\alpha^2 g_{\mu\nu} +2\alpha\beta
\,\mathcal{G}^\text{eff}_{\mu\nu} +\beta^2 f_{\mu\nu}\,,
\end{equation}
where $\alpha$ and $\beta$ are arbitrary constants, and we have
defined
\begin{equation}\label{Geff}
\mathcal{G}^\text{eff}_{\mu\nu} =
g_{\mu\rho}X^\rho_\nu=f_{\mu\rho}Y^\rho_\nu\,,
\end{equation}
where the square root matrix is defined by $X=\sqrt{g^{-1}f}$,
together with its inverse $Y=\sqrt{f^{-1}g}$. Interchanging the two
metrics $g$ and $f$ does not change the form of the
metric~\eqref{effmetric}--\eqref{Geff} except for a redefinition of
the parameters $\alpha$ and $\beta$. Notice that
$\mathcal{G}^\text{eff}_{\mu\nu}$ can be proved to be
automatically symmetric \cite{Baccetti:2012re, Hassan:2012wr}, and in
\cite{Blanchet:2015sra} it was shown, that it corresponds to the
composite metric considered in the previous model~\cite{BB14}. The
square root of the determinant of $g^\text{eff}_{\mu\nu}$ is given in
either forms respectively associated with the $g$ or $f$ metrics, by
\begin{equation}\label{detgeff}
\sqrt{-g_\text{eff}}=\sqrt{-g} \,\det\bigl(\alpha\Id +\beta
X\bigr)=\sqrt{-f} \,\det\bigl(\beta\Id +\alpha Y\bigr)\,.
\end{equation}
More explicitly, introducing the elementary symmetric polynomials
$e_n(X)$ and $e_n(Y)$ of the square root matrices $X$ or
$Y$,\footnote{As usual we denote the traces of rank-2 tensors as
  $X^{\mu}_{\mu}=[X]$, $X^{\mu}_{\nu} X^{\nu}_{\mu}=[X^2]$, and so
  on. Recall in particular that $\sqrt{-g}\,e_n(X) =
  \sqrt{-f}\,e_{4-n}(Y)$, with $e_4(X)=\text{det}(X)$ and
  $e_4(Y)=\text{det}(Y)$.}
{\allowdisplaybreaks 
\begin{subequations}\begin{eqnarray}\label{polynomials}
e_0(X)&=&1\,, \\ e_1(X)&=& \bigl[X\bigr]\,, \\ e_2(X)&=&
\frac{1}{2}\bigl(\bigl[X\bigr]^2-\bigl[X^2\bigr]\bigr)\,, \\ e_3(X)&=&
\frac{1}{6}\bigl(\bigl[X\bigr]^3-3\bigl[X\bigr]\bigl[X^2\bigr]
+2\bigl[X^3\bigr]\bigr)\,, \\ e_4(X)&=&
\frac{1}{24}\bigl(\bigl[X\bigr]^4-6\bigl[X\bigr]^2\bigl[X^2\bigr]
+3\bigl[X^2\bigr]^2 +
8\bigl[X\bigr]\bigl[X^3\bigr]-6\bigl[X^4\bigr]\bigr)\,,
\end{eqnarray}
\end{subequations}}
we can write, still in symmetric guise,
\begin{equation}
\sqrt{-g_\text{eff}} = \sqrt{-g} \sum_{n=0}^4\alpha^{4-n}\beta^{n}
e_n(X) = \sqrt{-f} \sum_{n=0}^4\alpha^{n}\beta^{4-n} e_n(Y)\,.
\end{equation}
In this form we see that~\eqref{detgeff} corresponds to the right form
of the acceptable potential interactions between the metrics $g$ and
$f$.

We can first vary the Lagrangian~\eqref{lagrangian} with respect to
the two metrics $g$ and $f$, which yields the following two covariant
Einstein field equations~\cite{Schmidt-May:2014xla}
\begin{subequations}\label{EFEgf}
\begin{align}
&M_g^2 \,Y^{(\mu}_\rho G_g^{\nu)\rho} - m^2 M_\text{eff}^2
  \sum_{n=0}^3\alpha^{4-n}\beta^{n} \,Y^{(\mu}_\rho U_{(n)}^{\nu)\rho}
  \nonumber\\ &\qquad\qquad\qquad\quad =
  Y^{(\mu}_\rho\Bigl(T_\text{bar}^{\nu)\rho}+T_g^{\nu)\rho}\Bigr) +
  \alpha \frac{\sqrt{-g_\text{eff}}}{\sqrt{-g}}\Bigl( \alpha
  Y^{(\mu}_\rho T_{g_{\text{eff}}}^{\nu)\rho}+\beta
  T_{g_{\text{eff}}}^{\mu\nu}\Bigr)\,,\\ &M_f^2 \,X^{(\mu}_\rho
  G_f^{\nu)\rho} - m^2 M_\text{eff}^2
  \sum_{n=0}^3\alpha^{n}\beta^{4-n} \,X^{(\mu}_\rho V_{(n)}^{\nu)\rho}
  \nonumber\\ &\qquad\qquad\qquad\quad = X^{(\mu}_\rho T_f^{\nu)\rho}
  + \beta \frac{\sqrt{-g_\text{eff}}}{\sqrt{-f}}\Bigl( \beta
  X^{(\mu}_\rho T_{g_{\text{eff}}}^{\nu)\rho}+\alpha
  T_{g_\text{eff}}^{\mu\nu}\Bigr)\,,
\end{align}
\end{subequations}
where $G_g^{\mu\nu}$ and $G_f^{\mu\nu}$ are the Einstein tensors for
the $g$ and $f$ metrics. The tensors $U_{(n)}^{\mu\nu}$ and
$V_{(n)}^{\mu\nu}$ are defined from the following sums of
matrices~\cite{Hassan:2011vm}\footnote{Here we adopt a slight change
  of notation with respect to~\cite{Hassan:2011vm}. Our notation is
  related to the one of~\cite{Hassan:2011vm} by $U_{(n)}=(-)^n
  Y_{(n)}(X)$ and $V_{(n)} = (-)^n Y_{(n)}(Y)$.}
\begin{subequations}
\begin{align}
&U_{(n)} = \sum_{p=0}^n (-)^p e_{n-p}(X) X^p\,,\\ &V_{(n)} =
  \sum_{p=0}^n (-)^p e_{n-p}(Y) Y^p\,,
\end{align}
\end{subequations}
by raising or lowering indices with their respective metrics, thus
$U_{(n)}^{\mu\nu}=g^{\mu\rho}U_{(n)\rho}^{\nu}$ and
$V_{(n)}^{\mu\nu}=f^{\mu\rho}V_{(n)\rho}^{\nu}$. By the same property
which makes~\eqref{Geff} to be symmetric, one can show that
$U_{(n)}^{\mu\nu}$ and $V_{(n)}^{\mu\nu}$ are indeed automatically
symmetric.

In the right sides of~\eqref{EFEgf} the stress-energy tensors
$T_\text{bar}^{\mu\nu}$ and $T_g^{\mu\nu}$ are defined with respect to
the metric $g$, $T_f^{\mu\nu}$ is defined with respect to $f$ and
$T_{g_{\text{eff}}}^{\mu\nu}$ with respect to $g_\text{eff}$. Thus,
for pressureless baryonic and dark matter fluids we have
$T_\text{bar}^{\mu\nu}=\rho_\text{bar} u_\text{bar}^\mu
u_\text{bar}^\nu$, $T_g^{\mu\nu}=\rho_g u_g^\mu u_g^\nu$ and
$T_f^{\mu\nu}=\rho_f u_f^\mu u_f^\nu$ in terms of the corresponding
scalar densities and normalized four velocities, while
$T_{g_{\text{eff}}}^{\mu\nu}$ represents in fact the stress-energy
tensor of the vector field $\mathcal{A}_\mu$, \textit{i.e.}
\begin{equation}\label{Tgeff}
T_{g_{\text{eff}}}^{\mu\nu} = M_\text{eff}^2\Bigl[\mathcal{W}_{\mathcal{X}}
  \,\mathcal{F}^{\mu\rho}\mathcal{F}^\nu_{\phantom{\nu}\rho} +
  \lambda\mathcal{W}\,g_\text{eff}^{\mu\nu}\Bigr]\,,
\end{equation}
where $\mathcal{W}_{\mathcal{X}}$ is the derivative of $\mathcal{W}$
with respect to its argument $\mathcal{X}$ defined by~\eqref{X}.

The equation of motion for the baryons follows a geodesic law
$a^\text{bar}_\mu \equiv u_\text{bar}^\nu \nabla^{g}_\nu
u^\text{bar}_\mu=0$, whereas the equations of motion for the dark
matter particles are non-geodesic,
\begin{subequations}\label{EOMdm}
\begin{align}
a^g_\mu &= r_g u_g^\nu \mathcal{F}_{\mu\nu}\,,\\ a^f_\mu &= - r_f
u_f^\nu \mathcal{F}_{\mu\nu}\,,
\end{align}
\end{subequations}
where the accelerations $a^g_\mu \equiv u_g^\nu \nabla^{g}_\nu
u^g_\mu$ and $a^f_\mu \equiv u_f^\nu \nabla^{f}_\nu u^f_\mu$ are
defined with their respective metrics. On the other hand, the equation
for the vector field yields
\begin{equation}\label{eqAmu}
\nabla^{g_\text{eff}}_\nu\Bigl[
  \mathcal{W}_{\mathcal{X}}\mathcal{F}^{\mu\nu}\Bigr] =
\frac{1}{M_\text{eff}^2}\bigl(j^\mu_g-j^\mu_f\bigr)\,,
\end{equation}
with $\nabla^{g_\text{eff}}_\nu$ denoting the covariant derivative
associated with $g_\text{eff}$. This equation is obviously compatible
with the conservation of the currents, since by~\eqref{currents} we
have also $\nabla^{g_\text{eff}}_\mu j_g^\mu=0$ and $\nabla^{g_\text{eff}}_\mu
j_f^\mu=0$. The equation~\eqref{eqAmu} can be also written as
\begin{equation}\label{divtaueff}
\nabla_{g_\text{eff}}^\nu T^{g_\text{eff}}_{\mu\nu} = -
\bigl(j^\nu_g-j^\nu_f \bigr)\mathcal{F}_{\mu\nu}\,.
\end{equation}
Combining the equations of motion~\eqref{EOMdm} with~\eqref{divtaueff}
we obtain the conservation law
\begin{equation}\label{conslaw}
\sqrt{-g}\nabla_g^\nu T^g_{\mu\nu} + \sqrt{-f}\nabla_f^\nu
T^f_{\mu\nu} + \sqrt{-g_\text{eff}}\nabla_{g_\text{eff}}^\nu
T^{g_\text{eff}}_{\mu\nu} = 0\,.
\end{equation}
Alternatively, such global conservation law can be obtained from
the scalarity of the total matter action under general
diffeomorphisms.

\section{Linear perturbations}\label{sec:linear_pert}

In this section we will be interested in computing the linear
perturbations of our model, which will be extensively used in the
following section for the study of the polarisation
mechanism. Following the ideas of~\cite{BB14} we will assume that the
two fluids of dark matter particles are slightly displaced from the
equilibrium configuration by displacement vectors $y_g^\mu$ and
$y_f^\mu$. This makes the dark matter medium to act as an analogue of
a relativistic plasma in electromagnetism. The dark matter currents
will be slightly displaced from the equilibrium current
$j^\mu_0=\rho_0 u_0^\mu$ as well, with $j^\mu_0$ satisfying
$\nabla^{g_{\text{eff}}}_\mu j^\mu_0=0$. Furthermore, we will perturb
the two metrics around a general background in the following
way\footnote{By which we mean, for instance, $g_{\mu\nu} =
  (\bar{g}_{\mu\rho} +
  h_{\mu\rho})\bar{g}^{\rho\sigma}(\bar{g}_{\sigma\nu} +
  h_{\sigma\nu})$.}
\begin{subequations}\label{perturbedFlat}
\begin{align}
g_{\mu\nu} &= (\bar{g}_{\mu\nu}+h_{\mu\nu})^2\,,\\ f_{\mu\nu} &=
(\bar{f}_{\mu\nu}+\ell_{\mu\nu})^2\,,
\end{align}\end{subequations}
with main interest in background solutions where
$\bar{g}_{\mu\nu}=\bar{f}_{\mu\nu}$. In the latter case the potential
interactions take the form
\begin{equation}\label{detgeffexpl}
\sqrt{-g_\text{eff}} = \sqrt{-\bar{g}}\sum_{n=0}^4
(\alpha+\beta)^{4-n} e_n(k)\,,
\end{equation}
where $k_{\mu\nu}=\alpha h_{\mu\nu} + \beta \ell_{\mu\nu}$, and
traces are defined with the common background metric
$\bar{g}=\bar{f}$.

Working at first order in the displacement of the dark matter
particles and assuming that their gradients are of the same order as
the metric perturbations, the dark matter currents can then be
expanded as~\cite{BB14}
\begin{subequations}\label{ansatz_currents}
\begin{align}
j_{g}^\mu &= j_0^\mu +2\nabla^{g_{\text{eff}}}_\nu (j_0^{[\nu}
  \!\!\perp^{\mu]}_\rho y^\rho_{g}) + \calO(2)\,,\\ j_{f}^\mu &=
j_0^\mu +2\nabla^{g_{\text{eff}}}_\nu (j_0^{[\nu}
  \!\!\perp^{\mu]}_\rho y^\rho_{f}) + \calO(2)\,,
\end{align}\end{subequations}
with the projector operator perpendicular to the four velocity of the
equilibrium $u_0^\mu$ given as
$\perp^{\mu\nu}=g_{\text{eff}}^{\mu\nu}+u_0^\mu u_0^\nu$, and the
remainder $\mathcal{O}(2)$ indicating that the expansion is at first
order. The desired plasma-like solution is achieved after plugging our
ansatz~\eqref{ansatz_currents} into the equations of the vector
field~\eqref{eqAmu}, which results in
\begin{equation}\label{WXFsol}
\mathcal{W}_{\mathcal{X}}\mathcal{F}^{\mu\nu} =
-\frac{2}{M_\text{eff}^2} \,j_0^{[\mu} \xi^{\nu]} + \calO(2)\,,
\end{equation}
where we have defined the projected relative displacement as
$\xi^\mu\equiv\perp^\mu_\nu\!(y^\nu_{g}-y^\nu_{f})$. Note that
  $\xi^\mu$ is necessarily a space-like vector. In
Sec.~\ref{sec:polarisation_mond} we shall see that the spatial
components $\xi^i$ of this vector, which can be called a dipole
moment, define in the non-relativistic limit the polarisation field of
the dark matter medium as $P^i=\rho^*_0\xi^i$, where $\rho^*_0$ is the
coordinate density associated with $\rho_0$. But for the moment we
only need to notice that $\xi^\mu$ is a first order quantity,
therefore the field strength $\mathcal{F}^{\mu\nu}$ is itself a first
order quantity, and hence the stress-energy tensor~\eqref{Tgeff} with
respect to $g_{\text{eff}}$ is already of second order,\footnote{We
  are anticipating the form of $\mathcal{W}$ given by~\eqref{WX} and
  which implies $\mathcal{W}=\calO(2)$ and
    $\mathcal{W}_{\mathcal{X}}=1+\calO(1)$.}
\begin{equation}\label{Teff0}
T_{g_{\text{eff}}}^{\mu\nu}=\calO(2)\,.
\end{equation}
For this reason, in the case of our desired plasma-like solution the
contributions of the self-interactions of the internal vector field
will be at least third order in perturbation at the level of the
action. For more detail see the comprehensive derivations
in~\cite{BB14}.

Let us first expand the Lagrangian~\eqref{lagrangian} to first order
in perturbation around a common background $\bar{g}=\bar{f}$. We
assume that there is no matter in the background, so the matter
interactions do not contribute at that
order. Using~\eqref{detgeffexpl} we obtain (modulo a total divergence)
\begin{equation}
\mathcal{L} = \mathcal{L}_{\bar{g}} + \sqrt{-\bar{g}}\Bigl[-
  \bigl(M_g^2 h^{\mu\nu} + M_f^2
  \ell^{\mu\nu}\bigr)G_{\mu\nu}^{\bar{g}} + m^2
  M_{\text{eff}}^2(\alpha+\beta)^3 \bigl[k\bigr] \Bigr] + \calO(2)\,,
\end{equation}
where $G_{\mu\nu}^{\bar{g}}$ is the Einstein tensor for the background
metric $\bar{g}$, and $\mathcal{L}_{\bar{g}}$ is the background value
of the Lagrangian. In order for the background $\bar{g}_{\mu\nu}$ in
our interest to be a solution of the theory, we have to impose that
linear perturbations vanish. We find
\begin{equation}\label{background}
G_{\mu\nu}^{\bar{g}} =
m^2\frac{M_{\text{eff}}^2}{M_g^2}\alpha(\alpha+\beta)^3\bar{g}_{\mu\nu}
=
m^2\frac{M_{\text{eff}}^2}{M_f^2}\beta(\alpha+\beta)^3\bar{g}_{\mu\nu}\,.
\end{equation}
This is only possible if
\begin{equation}\label{consistent}
\frac{\alpha}{M_g^2} = \frac{\beta}{M_f^2}\,.
\end{equation}
With this choice we guarantee that we are expanding around the
background $\bar{g}_{\mu\nu}=\bar{f}_{\mu\nu}$ that is a solution to
the background equations of motion. 

As next we shall compute the action to second order in
perturbations. Let us emphasize again that thanks to the plasma-like
ansatz~\eqref{ansatz_currents} the self-interaction of the internal
vector field does not contribute to this order, see~\eqref{Teff0}. Our
quadratic Lagrangian above the background $\bar{g}_{\mu\nu}$ simply
becomes
\begin{align}
\mathcal{L} = \mathcal{L}_{\bar{g}} &+ \sqrt{-\bar{g}}\biggl[-M_g^2
  h^{\rho\sigma} \bar{\mathcal{E}}^{\mu\nu}_{\rho\sigma} h_{\mu\nu}
  -M_f^2 \ell^{\rho\sigma} \bar{\mathcal{E}}^{\mu\nu}_{\rho\sigma}
  \ell_{\mu\nu} +
  h_{\mu\nu}\bigl(T^{\mu\nu}_\text{bar}+T^{\mu\nu}_g\bigr) +
  \ell_{\mu\nu}T^{\mu\nu}_f \nonumber\\ & \qquad\qquad + \frac{1}{2}
  m^2 M_{\text{eff}}^2(\alpha+\beta)^2
  \bigl(\bigl[k\bigr]^2-\bigl[k^2\bigr]\bigr) \biggr] + \calO(3)\,,
\end{align}
where $\bar{\mathcal{E}}^{\rho\sigma}_{\mu\nu}$ is the Lichnerowicz
operator on the background $\bar{g}$ given in the general case by
\begin{align}\label{lichne}
-2\bar{\mathcal{E}}_{\mu\nu}^{\rho\sigma}h_{\rho\sigma} &=
\Box_{\bar{g}}\bigl(h_{\mu\nu}-\bar{g}_{\mu\nu}h\bigr) +
\nabla^{\bar{g}}_\mu\nabla^{\bar{g}}_\nu h - 2 \nabla^{\bar{g}}_{(\mu}
H_{\nu)} + \bar{g}_{\mu\nu}\nabla^{\bar{g}}_{\rho} H^{\rho}
\nonumber\\ & - 2 C^{\bar{g}}_{\mu\rho\sigma\nu}h^{\rho\sigma} -
\frac{2}{3}R_{\bar{g}}\Bigl( h^{\mu\nu}-\frac14 \bar{g}_{\mu\nu}
h\Bigr)\,,
\end{align}
where $h=[h]=\bar{g}^{\mu\nu}h_{\mu\nu}$,
$H_\mu=\nabla_{\bar{g}}^{\nu}h_{\mu\nu}$, with
$C^{\bar{g}}_{\mu\rho\sigma\nu}$ denoting the Weyl curvature of the
background metric. Of course this expression is to be simplified using
the background equations~\eqref{background}. Also bear in mind that
$\alpha M_f^2=\beta M_g^2$ for having a common background
$\bar{g}=\bar{f}$. The field equations obtained by varying with
respect to $h_{\mu\nu}$ and $\ell_{\mu\nu}$ yield
\begin{subequations}\label{eomgh}
\begin{align}
& -2M_g^2\bar{\mathcal{E}}_{\rho\sigma}^{\mu\nu}h^{\rho\sigma} +
  T^{\mu\nu}_\text{bar} + T^{\mu\nu}_g + \alpha(\alpha+\beta)^2 m^2
  M_{\text{eff}}^2\Bigl(\bar{g}^{\mu\nu}k-k^{\mu\nu}\Bigr) =
  \calO(2)\,,\\ &
  -2M_f^2\bar{\mathcal{E}}_{\rho\sigma}^{\mu\nu}\ell^{\rho\sigma} +
  T^{\mu\nu}_f + \beta(\alpha+\beta)^2 m^2
  M_{\text{eff}}^2\Bigl(\bar{g}^{\mu\nu}k-k^{\mu\nu}\Bigr) =
  \calO(2)\,,
\end{align}\end{subequations}
where we recall that $k^{\mu\nu}=\alpha h^{\mu\nu}+ \beta
\ell^{\mu\nu}$ and $k=[k]=\bar{g}_{\mu\nu}k^{\mu\nu}$. Of course those
equations can also be recovered directly from the general Einstein
field equations~\eqref{EFEgf}.

In the present paper we shall mostly make use of that combination of
the two Einstein field equations~\eqref{eomgh} which corresponds to
the propagation of a massless spin-2 field. We subtract the two
equations from each other so as to cancel the mass term, resulting in
\begin{equation}\label{massless}
-2\bar{\mathcal{E}}_{\rho\sigma}^{\mu\nu}
\biggl(\frac{M_g^2}{\alpha}h^{\rho\sigma} -
\frac{M_f^2}{\beta}\ell^{\rho\sigma}\biggr) +
\frac{1}{\alpha}\left(T^{\mu\nu}_\text{bar}+T^{\mu\nu}_g\right) -
\frac{1}{\beta}T^{\mu\nu}_f = \calO(2)\,.
\end{equation}
However we shall also use the contribution of the mass term in the
linearised Bianchi identities associated with~\eqref{eomgh}. Thus,
last but not least we can act on~\eqref{eomgh} with
$\nabla_\nu^{\bar{g}}$, yielding
\begin{equation}\label{bianchi}
\nabla_\nu^{\bar{g}}\bigl(k^{\mu\nu} - \bar{g}^{\mu\nu}k\bigr) =
\calO(2)\,.
\end{equation}
In deriving this relation we use the result that, for instance,
$\nabla^{\bar{g}}_\nu T_g^{\mu\nu}=\rho_g a_g^\mu=\calO(2)$ which
comes from the equations of motion~\eqref{EOMdm} and the fact that
there is no matter in the background. We shall see in the next section
how important is this relation for the present model to recover the
looked-for phenomenology of dark matter at galactic scales.
%


\section{Polarisation mechanism and MOND}\label{sec:polarisation_mond}

We next consider the Newtonian or non-relativistic (NR) limit of our
model. In the previous section we derived the field equations at
linear order around a vacuum background metric $\bar{g}$, which
necessarily obeys the equations~\eqref{background}. Thus, the
background is a de Sitter solution, with cosmological constant given
by the graviton's mass, $\Lambda\sim m^2$. When computing the
Newtonian limit applied to describe the physics of the local universe,
\textit{e.g.}  the solar system or a galaxy at low redshift, we shall
neglect the cosmological constant and shall approximate the de Sitter
metric $\bar{g}$ by a flat Minkowski background $\eta$.

The best way to implement (and justify) this approximation is to
perform an expansion on small scales, say $r\to 0$. When we solve for
the gravitational field in the solar system or a galaxy embedded in
the de Sitter background, we get terms like $1 - 2M r^{-1} - \Lambda
r^2/3$ (de Sitter-Schwarzschild solution), where $M$ is the mass of
the Sun or the galaxy. On small scales the term $2M r^{-1}$ will
dominate over $\Lambda r^2/3$, so we can ignore the influence of the
cosmological constant if the size of the system is $\ll (6
M/\Lambda)^{1/3}$. In the present case we shall require that the size
of our system includes the very weak field region far from the system
where the MOND formula applies. In that case the relevant scale is
$r_0=\sqrt{M/a_0}$, where $a_0\simeq 1.2\times
10^{-10}\,\text{m}/\text{s}^2$ is the MOND acceleration. So our
approximation will make sense provided that $r_0\ll
(6M/\Lambda)^{1/3}$. For a galaxy with mass $M\sim 10^{11}\,M_\odot$
the MOND transition radius is $r_0\sim 10\,\text{kpc}$. With the value
of the cosmological constant $\Lambda\sim (3\,\text{Gpc})^{-2}$ we
find $(6M/\Lambda)^{1/3}\sim 700\,\text{kpc}$, so we are legitimate to
neglect the influence of the cosmological constant. For the solar
system, $r_0\sim 0.04\,\text{pc}$ while $(6M_\odot/\Lambda)^{1/3}\sim
150\,\text{pc}$ and the approximation is even better.

Actually we shall make the approximation $\bar{g}\simeq\eta$ only on
the particular massless spin-2 combination of the two metrics given
by~\eqref{massless}, namely
\begin{equation}\label{masslesscomb}
-2\bar{\mathcal{E}}_{\rho\sigma}^{\mu\nu}
\Bigl(\frac{M_g^2}{\alpha}h^{\rho\sigma} -
\frac{M_f^2}{\beta}\ell^{\rho\sigma}\Bigr) =
-\frac{1}{\alpha}\left(T^{\mu\nu}_\text{bar} +
T^{\mu\nu}_g\right)+\frac{1}{\beta} T^{\mu\nu}_f + \calO(2)\,,
\end{equation}
where the Lichnerowicz operator~\eqref{lichne} is now the flat one,
and also in the constraint equation~\eqref{bianchi}. We shall not need
to consider the other, independent (massive) combination of the two
metrics. In the NR limit when $c\to\infty$ we parametrize the two
metric perturbations $h_{\mu\nu}$ and $\ell_{\mu\nu}$ by single
Newtonian potentials $U_g$ and $U_f$ such that
\begin{subequations}\label{PNexp}
\begin{eqnarray}
&h_{00} = \displaystyle \frac{U_g}{c^2} +
  \calO\left(\frac{1}{c^4}\right)\,,\qquad &\ell_{00} =
  \frac{U_f}{c^2} +
  \calO\left(\frac{1}{c^4}\right)\,,\\ &\hspace{-1cm}h_{0i} =
  \displaystyle \calO\left(\frac{1}{c^3}\right)\,,\qquad &\ell_{0i} =
  \calO\left(\frac{1}{c^3}\right)\,,\\ &~~h_{ij} = \displaystyle
  \delta_{ij}\frac{U_g}{c^2} +
  \calO\left(\frac{1}{c^4}\right)\,,\qquad &\ell_{ij} =
  \delta_{ij}\frac{U_f}{c^2} + \calO\left(\frac{1}{c^4}\right)\,.
\end{eqnarray}
\end{subequations}
We shall especially pay attention to the potential $U_g$ since it
represents the ordinary Newtonian potential felt by ordinary baryonic
matter. Similarly we parametrize the internal vector field
$\mathcal{A}_\mu$ by a single Coulomb scalar potential $\phi$ such
that
\begin{subequations}
\begin{align}
\mathcal{A}_{0} &= \displaystyle \frac{\phi}{c^2} +
  \calO\left(\frac{1}{c^4}\right)\,,\\ \mathcal{A}_{i} &=
  \displaystyle \calO\left(\frac{1}{c^3}\right)\,.
\end{align}
\end{subequations}
It is now straightforward to show that~\eqref{masslesscomb} reduces in
the NR limit to a single scalar equation for a combination of the
Newtonian potentials $U_g$ and $U_f$,\footnote{We shall adopt the
  usual boldface notation for ordinary three-dimensional Euclidean
  vectors. Also, from now on we no longer write the neglected
  remainders $\calO(1/c^2)$.}
\begin{equation}\label{scalareq}
\Delta\left(\frac{2M_g^2}{\alpha}U_g - \frac{2M_f^2}{\beta}U_f\right)
= -\frac{1}{\alpha}\bigl(\rho^*_\text{bar} +
\rho^*_g\bigr)+\frac{1}{\beta} \rho^*_f\,,
\end{equation}
where $\Delta=\bm{\nabla}^2$ is the ordinary Laplacian, and
$\rho^*_\text{bar}$, $\rho^*_g$ and $\rho^*_f$ are the ordinary
Newtonian (coordinate) densities, satisfying usual continuity
equations such as $\partial_t\rho^*_g+\bm{\nabla}\cdot(\rho^*_g
\bm{v}_g)=0$. Notably, all non-linear corrections $\calO(2)$
in~\eqref{masslesscomb} are negligible. In principle, we should
\textit{a priori} allow some parametrized post-Newtonian (PPN)
coefficients in the spatial components $ij$ of the
metrics~\eqref{PNexp}, say $\gamma_g$ and $\gamma_f$. However, when
plugging~\eqref{PNexp} into~\eqref{masslesscomb} we determine from the
$ij$ components of~\eqref{masslesscomb} that in fact $\gamma_g=1$ and
$\gamma_f=1$. Similarly we find that the equations~\eqref{eqAmu}
governing the internal vector field reduce to a single Coulomb type
equation,
\begin{equation}\label{coulomb}
\bm{\nabla}\cdot\Bigl[\mathcal{W}_{\mathcal{X}}\bm{\nabla}\phi\Bigr] =
\frac{1}{M^2_\text{eff}}\bigl(r_g \rho^*_g - r_f \rho^*_f\bigr)\,.
\end{equation}
Notice that the constants $\alpha$ and $\beta$ in the effective metric
$g_\text{eff}$ to which is coupled the internal field intervene in the
NR limit only in the expression
\begin{equation}\label{XNR}
\mathcal{X} = \frac{\vert\bm{\nabla}\phi\vert^2}{2\lambda
  (\alpha+\beta)^4}\,.
\end{equation}
Next, the equations of the baryons which are geodesic in the metric
$g$ reduce to
\begin{equation}\label{baryons}
\frac{\ud \bm{v}_\text{bar}}{\ud t} = \bm{\nabla}U_g\,,
\end{equation}
while the equations of motion~\eqref{EOMdm} of the dark matter
particles become
\begin{subequations}\label{eomNR}
\begin{align}
\frac{\ud \bm{v}_g}{\ud t} &= \bm{\nabla}U_g + r_g
\bm{\nabla}\phi\,,\\ \frac{\ud \bm{v}_f}{\ud t} &= \bm{\nabla}U_f -
r_f \bm{\nabla}\phi\,.
\end{align}
\end{subequations}

At this stage the important point is to recall that the two sectors
associated with the metrics $g$ and $f$ do not evolve independently
but are linked together by the constraint equation~\eqref{bianchi}. We
now show that this constraint provides a mechanism of gravitational
polarisation in the NR limit and yields the MOND equation for the
potential $U_g$ felt by baryons. In flat space-time this constraint is
\begin{equation}\label{constr}
\partial_\nu \bigl( k^{\mu\nu} - \eta^{\mu\nu} k\bigr) = \calO(2)\,,
\end{equation}
where $k^{\mu\nu} = \alpha h^{\mu\nu} + \beta \ell^{\mu\nu}$ and $k =
\eta^{\rho\sigma} k_{\rho\sigma}$. Plugging~\eqref{PNexp}
into~\eqref{constr} we readily obtain a constraint on the
gravitational forces felt by the DM particles, namely
\begin{equation}\label{constrNR}
\bm{\nabla}\bigl(\alpha U_g + \beta U_f\bigr) = 0\,.
\end{equation}
Note that the constraint~\eqref{constrNR} comes from a combination
between the $00$ and $ij$ components of the
metrics~\eqref{PNexp}. Using~\eqref{constrNR} we express the equations
of motion~\eqref{eomNR} solely in terms of the potential $U_g$ ruling
the baryons,
\begin{subequations}\label{eomDM}
\begin{align}
\frac{\ud \bm{v}_g}{\ud t} &= \bm{\nabla}U_g + r_g \bm{\nabla}\phi
\,,\\ \frac{\ud \bm{v}_f}{\ud t} &= -
\frac{\alpha}{\beta}\bm{\nabla}U_g - r_f \bm{\nabla}\phi \,.
\end{align}
\end{subequations}
As we see, the gravitational to inertial mass ratio
$m_\text{g}/m_\text{i}$ of the $f$ particles when measured with
respect to the $g$ metric is $m_\text{g}/m_\text{i}=-\alpha/\beta$.

We look for explicit solutions of~\eqref{eomDM} in the form of
plasma-like oscillations around some equilibrium solution. We
immediately see from~\eqref{eomDM} the possibility of an equilibrium
(\textit{i.e.} for which $\ud \bm{v}_g/\ud t=\ud \bm{v}_f/\ud t=0$)
when we have the following relation between constants,
\begin{equation}\label{relconst}
\frac{\alpha}{\beta} = \frac{r_f}{r_g} \,.
\end{equation}
The equilibrium holds when $\bm{\nabla}U_g + r_g \bm{\nabla}\phi=0$,
\textit{i.e.} when the Coulomb force annihilates the gravitational
force. To describe in the proper way the two DM fluids near or at
equilibrium we use the NR limits of the
relations~\eqref{ansatz_currents}, with an appropriate choice of the
equilibrium configuration. From~\eqref{eomDM} we note that
$(\alpha+\beta)\bm{v}_0=\alpha\bm{v}_g+\beta\bm{v}_f$ is constant,
hence we define the equilibrium in such a way that the two
displacement vectors $\bm{y}_g$ and $\bm{y}_f$ with respect to that
equilibrium obey $\alpha\bm{y}_g+\beta\bm{y}_f = \bm{0}$, and in
particular we choose the equilibrium velocity to be
$\bm{v}_0=\bm{0}$. We shall now define the relative displacement or
dipole moment vector by $\bm{\xi}=\bm{y}_g-\bm{y}_f$.\footnote{Here
  $\bm{y}_g$ and $\bm{y}_f$ denote the spatial components of the
  displacement four vectors $\perp^\mu_\nu\!y^\nu_{g}$ and
  $\perp^\mu_\nu\!y^\nu_{f}$ considered in Sec.~\ref{sec:linear_pert},
  see~\eqref{ansatz_currents}--\eqref{WXFsol}.}
Hence~\eqref{ansatz_currents} imply
\begin{subequations}\label{relpolar}
\begin{align}
r_g \rho^*_g &= \rho^*_0 -
\frac{\beta}{\alpha+\beta}\bm{\nabla}\cdot\bm{P}\,,\\ r_f \rho^*_f &=
\rho^*_0 + \frac{\alpha}{\alpha+\beta}\bm{\nabla}\cdot\bm{P}\,,
\end{align}\end{subequations}
where $\rho^*_0$ is the common density of the two DM fluids at
equilibrium in the absence of external perturbations (far from any
external mass), and $\bm{P}=\rho^*_0\bm{\xi}$ is the polarisation. The
velocity fields of the two fluids are
\begin{equation}\label{exprvel}
\bm{v}_g=\frac{\beta}{\alpha+\beta}\frac{\ud\bm{\xi}}{\ud
  t}\,,\qquad\bm{v}_f=-\frac{\alpha}{\alpha+\beta}\frac{\ud\bm{\xi}}{\ud
  t}\,,
\end{equation}
where $\ud \bm{\xi}/\ud
t=\partial_t\bm{\xi}+\bm{v}_0\cdot\bm{\nabla}\bm{\xi}$ denotes the
convective derivative, which reduces here to the ordinary derivative
since $\bm{v}_0=\bm{0}$.

By inserting~\eqref{relpolar} into~\eqref{coulomb} we can solve for
the polarisation resulting in
\begin{equation}\label{solve}
\bm{P} = - M^2_\text{eff}\,\mathcal{W}_{\mathcal{X}}\bm{\nabla}\phi\,.
\end{equation}
Furthermore, by combining~\eqref{exprvel} and~\eqref{eomDM} and making
use of the previous solution for the polarisation~\eqref{solve}, we
readily arrive at a simple harmonic oscillator describing plasma like
oscillations around equilibrium, namely
\begin{equation}\label{harmosc}
\frac{\ud^2 \bm{\xi}}{\ud t^2} + \omega_g^2 \bm{\xi} =
\frac{\alpha+\beta}{\beta}\bm{\nabla}U_g\,.
\end{equation} 
In this case the plasma frequency is given by
\begin{equation}\label{plasmafreq}
\omega_g = \sqrt{\frac{\alpha+\beta}{\beta}
  \frac{r_g\,\rho^*_0}{M^2_\text{eff}\,\mathcal{W}_{\mathcal{X}}}}\,.
\end{equation} 

Next, consider the equation for the Newtonian potential $U_g$ in the
ordinary sector. Combining~\eqref{constrNR} with~\eqref{scalareq} we
readily obtain
\begin{equation}\label{poissonNR}
\Delta U_g =
-\frac{1}{2\left(M_g^2+\frac{\alpha^2}{\beta^2}M_f^2\right)}
\Bigl[\rho^*_\text{bar} + \rho^*_g - \frac{\alpha}{\beta}
  \rho^*_f\Bigr] \,.
\end{equation}
To recover the usual Newtonian limit we must impose, in geometrical
units,
\begin{equation}\label{limitN}
M_g^2 + \frac{\alpha^2}{\beta^2} M_f^2 = \frac{1}{8\pi}\,.
\end{equation}
This condition being satisfied, the right-hand side
of~\eqref{poissonNR} can be rewritten with the help
of~\eqref{relpolar} and the relation~\eqref{relconst}. We obtain an
ordinary Poisson equation but modified by polarisation effects,
\begin{equation}\label{poissonpol}
\bm{\nabla}\cdot\biggl[\bm{\nabla} U_g - 4\pi
  \frac{\bm{P}}{r_g}\biggr] = - 4 \pi \rho^*_\text{bar} \,.
\end{equation}
We still have to show that the polarisation will be aligned with the
gravitational field $\bm{\nabla}U_g$, \textit{i.e.} we grasp a
mechanism of gravitational polarisation. This is a consequence of the
constitutive relation~\eqref{solve} taken at the equilibrium point,
neglecting plasma like oscillations. At this point we have
$\bm{\nabla}U_g + r_g \bm{\nabla}\phi=0$. Thus,
\begin{equation}\label{solveeq}
\bm{P} =
\frac{M^2_\text{eff}}{r_g}\,\mathcal{W}_{\mathcal{X}}\,\bm{\nabla}U_g
\,.
\end{equation}
Finally~\eqref{poissonpol} with~\eqref{solveeq} takes exactly the form
of the Bekenstein-Milgrom equation~\cite{BekM84}. To recover the
correct deep MOND regime we must impose that when $\mathcal{X}\to 0$
\begin{equation}\label{deepmond}
1 - 4\pi\frac{M^2_\text{eff}}{r_g^2}\,\mathcal{W}_{\mathcal{X}} =
\frac{\vert\bm{\nabla}U_g\vert}{a_0} \,,
\end{equation}
where $a_0$ is the MOND acceleration scale. This is easily achieved
with the choice
\begin{equation}\label{WX}
\mathcal{W}(\mathcal{X}) = \mathcal{X} -
\frac{2}{3}(\alpha+\beta)^2\mathcal{X}^{3/2} +
\calO\left(\mathcal{X}^2\right) \,,
\end{equation}
together with fixing the constants $M_\text{eff}$ and $\lambda$ to the
values
\begin{equation}\label{determ}
M^2_\text{eff} = \frac{r_g^2}{4\pi}\,,\qquad \lambda =
\frac{a_0^2}{2}\,.
\end{equation}

We thus recovered gravitational polarisation and the MOND equation (in
agreement with the dielectric analogy of MOND~\cite{B07mond}) as a
natural consequence of bimetric gravity since it is made possible by
the constraint equation~\eqref{constr} linking together the two
metrics of bigravity.\footnote{In the previous model~\cite{BB14} we
  had to assume that some coupling constant $\varepsilon$ in the
  action tends to zero. With bimetric gravity and its nice potential
  interactions we see that no particular assumption is required.} The
polarisation mechanism is possible only if we can annihilate the
gravitational force by some internal force, here chosen to be a vector
field, and therefore assume a coupling between the two species of DM
particles living in the $g$ and $f$ sectors. 

Let us recapitulate the various constraints we have found on the
parameters in the original action~\eqref{lagrangian}. These are given
by~\eqref{relconst} for having a polarisation process,~\eqref{limitN}
for recovering the Poisson equation and~\eqref{determ} for having the
correct deep MOND regime. In addition, we recall~\eqref{consistent}
which was imposed in order to be able to expand the two metrics around
the same background. Strictly speaking the relation~\eqref{consistent}
is not necessary for the present calculation because we neglected the
influence of the background, and~\eqref{consistent} may be relaxed for
some applications. Finally, we note that the result can be simplified
by absorbing the remaining constant $r_g$ together with the sum
$\alpha+\beta$ into the following redefinitions of the vector field
and the effective metric: $\mathcal{A}_\mu\rightarrow
r_g\mathcal{A}_\mu$ and $g_{\mu\nu}^\text{eff}\rightarrow
(\alpha+\beta)^{-2}g_{\mu\nu}^\text{eff}$. When this is done, and
after redefining also the graviton mass $m$, we see that we can always
choose $r_g=1$ and $\alpha+\beta=1$ without loss of generality.

Finally the fully reduced form of the action~\eqref{lagrangian} reads
\begin{align}
\mathcal{L}_\text{final} &= \sqrt{-g}\biggl(\frac{M_g^2}{2}R_g -
\rho_\text{bar}-\rho_g\Bigl[ 1 - \mathcal{A}_\mu u_g^\mu\Bigr]\biggr)
+\sqrt{-f}\biggl(\frac{M_f^2}{2}R_f-\rho_f\Bigl[ 1 +
\frac{\alpha}{\beta}\mathcal{A}_\mu u_f^\mu\Bigr]\biggr)
\nonumber\\ &+\sqrt{-g_\text{eff}} \biggl[ \frac{m^2}{4\pi} +
  \frac{a_0^2}{8\pi}\,\mathcal{W}\bigl(\mathcal{X}\bigr)
  \biggr]\,,\label{lagreduced}
\end{align}
where we have moved for convenience the mass currents $J_g^\mu=\rho_g
u_g^\mu$ and $J_f^\mu=\frac{\alpha}{\beta}\rho_f u_f^\mu$ to the $g$
and $f$ sectors, where the effective metric $g^\text{eff}_{\mu\nu}$ of
bigravity is given by~\eqref{effmetric} but with $\alpha+\beta=1$, and
where the kinetic term of the vector field is defined by
\begin{subequations}
\begin{align}
\mathcal{X} &= -
\frac{\mathcal{F}^{\mu\nu}\mathcal{F}_{\mu\nu}}{2a_0^2}\,,\\ \mathcal{W}(\mathcal{X})
&= \mathcal{X} - \frac{2}{3}\mathcal{X}^{3/2} +
\calO\left(\mathcal{X}^2\right)\,.
\end{align}
\end{subequations}
In addition the coupling constants $M_g^2$ and $M_f^2$ are constrained
by~\eqref{limitN}. We can still further impose~\eqref{consistent} if
we insist that the two metrics $g$ and $f$ can be expanded around a
common background $\bar{g}=\bar{f}$.\footnote{A simple choice
  satisfying all these requirement is $\alpha=\beta=\frac{1}{2}$ and
  $M_g^2 = M_f^2 = \frac{1}{16\pi}$, \textit{i.e.} each coupling
  constant takes half the GR value. With this choice the plasma
  frequency~\eqref{plasmafreq} becomes
$$\omega_g = \sqrt{\frac{8\pi
        \rho^*_0}{\mathcal{W}_{\mathcal{X}}}}\,,$$ in agreement with
    the finding of the previous model~\cite{BB14}.}
Finally the theory depends also on the graviton's mass $m$, hopefully
to be related to the observed cosmological constant $\Lambda$, and the
MOND acceleration scale $a_0$. Note that in such unified approach
between the cosmological constant and MOND, it is natural to expect
that $a_0$ and $\Lambda$ should have comparable orders of magnitudes,
\textit{i.e.}  $a_0\sim\sqrt{\Lambda}$ which happens to be in very
good agreement with observations.


\section{The decoupling limit}\label{sec:decoupling}

In this section we would like to focus on the matter sector and
address the question whether or not the interactions between the
matter fields reintroduces the BD ghost~\cite{Boulware:1973my}. For
this we derive the decoupling limit and pay special attention to the
helicity-0 mode. Before doing so, we first restore the broken gauge
invariance by introducing the Stueckelberg fields in the $f$ metric
(with $a, b=0,1,2,3$)
\begin{equation}
f_{\mu\nu} \to \tilde f_{\mu\nu} = f_{ab}\partial_\mu \phi^a
\partial_\nu \phi^b\,,
\end{equation}
where the Stueckelberg fields can be further decomposed into their
helicity-0 $\pi$ and helicity-1 $A^a$ counterparts,
\begin{equation}
\phi^a=x^a-\frac{m A^a}{\Lambda^3_3} -\frac{f^{ab}\partial_b
  \pi}{\Lambda^3_3}\,.
\end{equation}
As next we can take the decoupling limit by sending $M_{g}, M_{f} \to
\infty$, while keeping the scale $\Lambda^3_3=M_g m^2$ constant. For
our purpose, it is enough to keep track of the contributions of the
helicity-0 mode $\pi$ to the matter interactions. The BD ghost is
hidden behind the higher derivative terms of the helicity-0
interactions after using all of the covariant equations of motion and
constraints. We will therefore neglect the contribution of the
helicity-1 field and assume $g_{\mu\nu}=\eta_{\mu\nu}$ and
\begin{equation}
\label{eq:fDL}
f_{\mu\nu} = \eta_{\mu\nu} \to \tilde f_{\mu\nu} =
\left(\eta_{\mu\nu}-\Pi_{\mu\nu}\right)^2\,,
\end{equation}
where $\Pi_{\mu\nu}$ stands for $\Pi_{\mu\nu} \equiv \partial_\mu
\partial_\nu \pi/\Lambda^3_3$. In the remaining of this section all
indices are raised and lowered with respect to the Minkowski metric
$\eta_{\mu\nu}$. The effective metric in the decoupling limit
corresponds to
\begin{equation}
g_{\mu\nu}^{\text{eff}} \to \tilde{g}_{\mu\nu}^{\text{eff}} = \bigl[
  (\alpha+\beta)\eta_{\mu\nu}-\beta \Pi_{\mu\nu} \bigr]^2\,.
\end{equation}
In~\cite{Blanchet:2015sra} the required criteria for ghost freedom in
the kinetic and potential terms were investigated in detail and hence
the new model was constructed in such a way that it fulfils these
criteria. In the decoupling limit, the contribution to the equation of
motion for the helicity-0 field coming from the potential term is at
most second order in derivative for the allowed potentials. Therefore,
we can concentrate on the problematic terms coming from the matter
interactions. Let us for instance consider the coupling between
$j^\mu_{\tilde f}$ and $\mathcal{A}_\mu$, thus\footnote{For simplicity
  we will drop the constants $\lambda$, $r_f$ and $M_{\text{eff}}$ and
  assume that $\mathcal{W}(\mathcal{X})=\mathcal{X}$ in the
  following.}
\begin{equation}
\mathcal{L}_\text{mat}=-\sqrt{-\tilde{f}}\rho_{{\tilde
    f}}-\sqrt{-\tilde{g}_{\text{eff}}}\,\mathcal{A}_\mu j^\mu_{\tilde
  f}+\frac{1}{4}\sqrt{-\tilde{g}_{\text{eff}}}\,\mathcal{F}_{\mu\nu}^2\,,
\end{equation}
which we can also write as 
\begin{equation}
\mathcal{L}_\text{mat}=-\sqrt{-\tilde{f}}\,\rho_{\tilde
  f}\bigl(1+\mathcal{A}_\mu u^\mu_{\tilde
  f}\bigr)+\frac{1}{4}\sqrt{-\tilde{g}_{\text{eff}}}
\,\mathcal{F}_{\mu\nu}^2\,.
\end{equation}
Let us first compute the equation of motion with respect to the vector
field, which is simply given by
\begin{equation}
\frac{\delta \mathcal{L}_\text{mat}}{\delta
  \mathcal{A}_\mu}=-\sqrt{\tilde{g}_{\text{eff}}}
\,\nabla^{\tilde{g}_{\text{eff}}}_\nu
\mathcal{F}^{\mu\nu}-\sqrt{-{\tilde f}}\rho_{\tilde f}u^\mu_{\tilde
  f}\,,
\end{equation}
which can be also expressed as
\begin{equation}\label{eomVF}
\nabla^{\tilde{g}_{\text{eff}}}_\nu
\mathcal{F}^{\mu\nu}=-\frac{\sqrt{-{\tilde
      f}}}{\sqrt{-\tilde{g}_{\text{eff}}}}\,\rho_{{\tilde f}}
u^\mu_{\tilde f}=-j^\mu_{\tilde f}\,.
\end{equation}
As next we can compute the contribution of the matter to the equation
of motion with respect to the helicity-0 field,
\begin{eqnarray}
\frac{\delta \mathcal{L}_\text{mat}}{\delta \pi}&=&\frac{\partial_\mu
  \partial_\nu}{\Lambda^3_3}\left( \frac{\delta
  \mathcal{L}_\text{mat}}{\delta {\tilde f}_{\rho\sigma}} \frac{\delta
  {\tilde f}_{\rho\sigma}}{\delta \Pi_{\mu\nu}} \right) +
\frac{\partial_\mu \partial_\nu}{\Lambda^3_3}\left( \frac{\delta
  \mathcal{L}_\text{mat}}{\delta \tilde{g}^{\text{eff}}_{\rho\sigma}}
\frac{\delta \tilde{g}^{\text{eff}}_{\rho\sigma}}{\delta \Pi_{\mu\nu}}
\right) \nonumber\\ &=& -\frac{\partial_\mu \partial_\nu}{\Lambda^3_3}
\left( \sqrt{-{\tilde f}}\,T^{\mu\rho}_{\tilde
  f}(\delta^\nu_\rho-\Pi^\nu_\rho)+\beta
\sqrt{-\tilde{g}_{\text{eff}}}\,T^{\mu\rho}_{\tilde{g}_{\text{eff}}}
\bigl[(\alpha+\beta)\delta^\nu_\rho-\beta\Pi^\nu_\rho\bigr]\right)\,,
\end{eqnarray}
where we made use of
\begin{eqnarray}
T^{\mu\nu}_{\tilde f}=\frac{2}{\sqrt{-{\tilde f}}} \frac{\delta
  \mathcal{L}_\text{mat}}{\delta \tilde{f}_{\mu\nu}}\,, \qquad
\text{and} \qquad
T^{\mu\nu}_{\tilde{g}_{\text{eff}}}=\frac{2}{\sqrt{-\tilde{g}_{\text{eff}}}}
\frac{\delta \mathcal{L}_\text{mat}}{\delta
  \tilde{g}^{\text{eff}}_{\mu\nu}}\,.
\end{eqnarray}
Applying one of the derivatives we have
\begin{eqnarray}
\frac{\delta \mathcal{L}_\text{mat}}{\delta \pi}&=& -\frac{
  \partial_\nu}{\Lambda^3_3} \left\{ \partial_\mu\bigl( \sqrt{-{\tilde
    f}}\,T^{\mu\rho}_{\tilde
  f}\bigr)(\delta^\nu_\rho-\Pi^\nu_\rho)-\sqrt{-{\tilde
    f}}\,T^{\mu\rho}_{\tilde f} \partial_\mu \Pi^\nu_\rho
\right. \nonumber\\ &&+\left. \beta \partial_\mu(
\sqrt{-\tilde{g}_{\text{eff}}}T^{\mu\rho}_{\text{eff}})
\bigl[(\alpha+\beta)\delta^\nu_\rho-\beta\Pi^\nu_\rho\bigr]
-\beta^2\sqrt{-\tilde{g}_{\text{eff}}}\,T^{\mu\rho}_{\tilde{g}_{\text{eff}}}\,
\partial_\mu \Pi^\nu_\rho\right\}\,.
\end{eqnarray}
The equation of motion for the vector field~\eqref{eomVF} can also be
written as
\begin{equation}
\nabla_{\tilde{g}_{\text{eff}}}^\mu T^{\tilde{g}_{\text{eff}}}_{\rho\mu} =
j_{\tilde{f}}^\mu \,\mathcal{F}_{\rho\mu}\,,
\end{equation}
which we can use to express
\begin{equation}
\nabla^{\tilde{g}_{\text{eff}}}_\mu
T^{\mu\rho}_{\tilde{g}_{\text{eff}}} =
\frac{1}{\sqrt{-\tilde{g}_{\text{eff}}}} \partial_\mu\left(
\sqrt{-\tilde{g}_{\text{eff}}}T^{\mu\rho}_{\tilde{g}_{\text{eff}}}\right)
+
\Gamma_{\mu\sigma}^{\rho\,\tilde{g}_{\text{eff}}}\,T^{\mu\sigma}_{\tilde{g}_{\text{eff}}}
= j_{\tilde f}^\mu \,\tilde{g}_{\text{eff}}^{\rho
  \kappa}\,\mathcal{F}_{\kappa\mu}\,,
\end{equation}
hence we have 
\begin{equation}
\partial_\mu \left(
\sqrt{-\tilde{g}_{\text{eff}}}T^{\mu\rho}_{\tilde{g}_{\text{eff}}}\right)
=\sqrt{-\tilde{g}_{\text{eff}}}\,j_{\tilde f}^\mu
\,\tilde{g}_{\text{eff}}^{\rho\kappa}\,\mathcal{F}_{\kappa\mu}
-\sqrt{-\tilde{g}_{\text{eff}}}\,\Gamma_{\mu\sigma}^{\rho\,\tilde{g}_{\text{eff}}}
T^{\mu\sigma}_{\tilde{g}_{\text{eff}}}\,.
\end{equation}
Furthermore, the conservation law [see~\eqref{conslaw}] gives
\begin{equation}
\sqrt{-\tilde{g}_{\text{eff}}} \nabla^{\tilde{g}_{\text{eff}}}_\mu
T^{\mu\rho}_{\tilde{g}_{\text{eff}}}+\sqrt{-\tilde{f}}
\,\nabla_\mu^{\tilde{f}} T^{\mu\rho}_{\tilde f}=0\,.
\end{equation}
Thus, we have further
\begin{equation}
\partial_\mu \left( \sqrt{-\tilde{f}}T^{\mu\rho}_{\tilde
  f}\right)=-\sqrt{-\tilde{g}_{\text{eff}}}j^\mu_{\tilde f}
\tilde{f}^{\rho\kappa}\mathcal{F}_{\kappa\mu}
-\sqrt{-\tilde{f}}\,\Gamma_{\mu\sigma}^{\rho\,\tilde{f}}T^{\mu\sigma}_{\tilde
  f}\,.
\end{equation}
Using the equation of motion for the vector field and the conservation
equation, the equation for the helicity-0 becomes
\begin{align}
\frac{\delta \mathcal{L}_\text{mat}}{\delta \pi} &= -\frac{
  \partial_\nu}{\Lambda^3_3} \left\{ \left(
-\sqrt{-\tilde{g}_{\text{eff}}}\,j^\mu_{\tilde{f}}
\tilde{f}^{\rho\kappa}\mathcal{F}_{\kappa\mu}-\sqrt{-\tilde{f}}
\,\Gamma_{\mu\sigma}^{\rho\,\tilde{f}}T^{\mu\sigma}_{\tilde f}
\right)(\delta^\nu_\rho-\Pi^\nu_\rho)-\sqrt{-\tilde{f}}\,T^{\mu\rho}_{\tilde
  f}\partial_\mu \Pi^\nu_\rho \right. \\ &+\left. \beta
\left(\sqrt{-\tilde{g}_{\text{eff}}} \,j_{\tilde f}^\mu
\tilde{g}_{\text{eff}}^{\rho\kappa}\mathcal{F}_{\kappa\mu}
-\sqrt{-\tilde{g}_{\text{eff}}}
\,\Gamma_{\mu\sigma}^{\rho\,\tilde{g}_{\text{eff}}}\,T^{\mu\sigma}_{\tilde{g}_{\text{eff}}}
\right) \bigl[(\alpha+\beta)\delta^\nu_\rho-\beta\Pi^\nu_\rho\bigr] -
\beta^2\sqrt{-\tilde{g}_{\text{eff}}}
\,T^{\mu\rho}_{\tilde{g}_{\text{eff}}}\partial_\mu
\Pi^\nu_\rho\right\} \,, \nonumber
\end{align}
which we can rewrite as
\begin{eqnarray}
\frac{\delta \mathcal{L}_\text{mat}}{\delta \pi}&=& -\frac{
  \partial_\nu}{\Lambda^3_3} \left\{
-\sqrt{-\tilde{g}_{\text{eff}}}j^\mu_{\tilde f}
\tilde{f}^{\rho\kappa}\mathcal{F}_{\kappa\mu}(\delta^\nu_\rho-\Pi^\nu_\rho)
-\sqrt{-\tilde{f}}T^{\mu\sigma}_{\tilde f}\mathcal{R}_{\mu\sigma}^\nu
\right. \nonumber\\ &&+\left. \beta \sqrt{-\tilde{g}_{\text{eff}}}
j_{\tilde f}^\mu
\tilde{g}_{\text{eff}}^{\rho\kappa}\mathcal{F}_{\kappa\mu}
\bigl[(\alpha+\beta)\delta^\nu_\rho-\beta\Pi^\nu_\rho\bigr]
-\beta^2\sqrt{-\tilde{g}_{\text{eff}}}T^{\mu\sigma}_{\tilde{g}_{\text{eff}}}
\tilde{\mathcal{R}}_{\mu\sigma}^{\nu}\right\}\,,
\end{eqnarray}
where we have introduced
\begin{subequations}\begin{align}
\mathcal{R}^{\nu}_{\mu\sigma} &=
\Gamma^{\rho\,\tilde{f}}_{\mu\sigma}(\delta^\nu_\rho-\Pi^\nu_\rho)
+\partial^\nu\Pi_{\mu\sigma}\,,\\ \tilde{\mathcal{R}}^{\nu}_{\mu\sigma}
&=
\Gamma^{\rho\,\tilde{g}_{\text{eff}}}_{\mu\sigma}\bigl[(\alpha+\beta)
  \delta^\nu_\rho-\Pi^\nu_\rho\bigr]+\beta^2\partial^\nu\Pi_{\mu\sigma}\,.
\end{align}
\end{subequations}
Using the fact that the Christoffel symbol with respect to the
$\tilde{f}$ metric is given by
\begin{equation}
\Gamma^{\rho\,\tilde{f}}_{\mu\sigma} =
-\tilde{f}^{\rho\kappa}(\delta^\nu_\kappa -\Pi^\nu_\kappa)\partial_\nu
\Pi_{\mu\sigma}\,,
\end{equation}
and further taking into account the following relation,
\begin{equation}\label{usefulRel}
(\delta^\mu_\rho-\Pi^\mu_\rho)\tilde{f}^{\rho\sigma}(\delta^\nu_\sigma
  -\Pi^\nu_\sigma)=\eta^{\mu\nu}\,,
\end{equation}
we see immediately that $\mathcal{R}^{\nu}_{\mu\sigma}=0$. Similarly,
using the definition of the Christoffel symbol with respect to the
$\tilde{g}_{\text{eff}}$ metric,
\begin{equation}
\Gamma^{\rho\,\tilde{g}_{\text{eff}}}_{\mu\sigma}
=-\tilde{g}_{\text{eff}}^{\rho\kappa}\bigl[(\alpha+\beta)\delta^\nu_\kappa
  -\beta\Pi^\nu_\kappa\bigr]\partial_\nu \Pi_{\mu\sigma}\,,
\end{equation}
and the relation 
\begin{equation}
\bigl[(\alpha+\beta)\delta^\mu_\rho-\beta\Pi^\mu_\rho\bigr]
\tilde{g}_{\text{eff}}^{\rho\sigma}\bigl[(\alpha+\beta)\delta^\nu_\sigma
  -\beta\Pi^\nu_\sigma\bigr]=\eta^{\mu\nu}\,,
\end{equation}
we can show that also $\tilde{\mathcal{R}}^{\nu}_{\mu\sigma}=0$. Thus,
the contribution of the matter part to the equation of motion for the
helicity-0 mode simplifies to
\begin{equation}
\frac{\delta \mathcal{L}_\text{mat}}{\delta \pi} = -\frac{
  \partial_\nu}{\Lambda^3_3} \left\{
-\sqrt{-\tilde{g}_{\text{eff}}}\,j^\mu_{\tilde f}
\,\tilde{f}^{\rho\kappa}\,\mathcal{F}_{\kappa\mu}(\delta^\nu_\rho-\Pi^\nu_\rho)
+ \beta \sqrt{-\tilde{g}_{\text{eff}}} \,j_{\tilde f}^\mu
\,\tilde{g}_{\text{eff}}^{\rho\kappa}\,\mathcal{F}_{\kappa\mu}
\bigl[(\alpha+\beta)\delta^\nu_\rho-\beta\Pi^\nu_\rho\bigr]\right\}
\,.
\end{equation}
Using~\eqref{usefulRel} this can be equally written as
\begin{eqnarray}
\frac{\delta \mathcal{L}_\text{mat}}{\delta \pi}&=& -\frac{
  \partial_\nu}{\Lambda^3_3} \left\{ -j^\mu_*
\mathcal{F}_{\kappa\mu}(\eta^{\nu\kappa}-\Pi^{\nu\kappa})^{-1} + \beta
j_*^\mu \mathcal{F}_{\kappa\mu}
\bigl[(\alpha+\beta)\eta^{\nu\kappa}-\beta\Pi^{\nu\kappa}\bigr]^{-1}\right\}\,,
\end{eqnarray}
where $j_{*}^\mu=\sqrt{-\tilde{g}_{\text{eff}}}\,j_{\tilde f}^\mu$ is
the coordinate current and represents the matter degrees of freedom
independent of the metric, while $\mathcal{F}_{\kappa\mu}$ is also
independent of the metric. We can now also apply the remaining
derivative in front,
\begin{eqnarray}
\frac{\delta \mathcal{L}_\text{mat}}{\delta \pi}&=& -\frac{1
}{\Lambda^3_3} \left\{ - \partial_\nu( j_{*}^\mu
\tilde{f}^{\rho\kappa}\mathcal{F}_{\kappa\mu})(\delta^\nu_\rho-\Pi^\nu_\rho)
+\frac{1
}{\Lambda^3_3}j_{*}^\mu
\tilde{f}^{\rho\kappa}\mathcal{F}_{\kappa\mu}\partial_\rho\Box\pi
\right. \nonumber\\ &&\left.\quad\qquad + \beta\partial_\nu(j_{*}^\mu
\tilde{g}_{\text{eff}}^{\rho\kappa}\mathcal{F}_{\kappa\mu})
\bigl[(\alpha+\beta)\delta^\nu_\rho-\beta\Pi^\nu_\rho\bigr]-\frac{1
}{\Lambda^3_3}\beta^2j_{*}^\mu
\tilde{g}_{\text{eff}}^{\rho\kappa}\mathcal{F}_{\kappa\mu}\partial_\rho\Box\pi
\right\}\,.
\end{eqnarray}
The contribution of the matter interactions to the equation of motion
for the $\pi$ field contains higher derivative terms which we are not
able to remove by invoking the covariant equations of motion for the
vector field and the dark matter particle. Hence, this reflects the
presence of a ghostly degree of freedom in the decoupling limit
through the matter interaction term. The scale $\Lambda_3$ is not the
cut-off scale of this theory anymore, but rather given by the scale of
the mass of the introduced ghost $m_\text{BD}$. For the precise
plasma-like background solution that we considered here, the
contributions of the matter interactions and the internal vector field
will be at least third order in perturbation at the level of the
action. Hence the ghost would enter at non-linear order in
perturbations, and does not show up in our linear perturbation
analysis of the gravitational polarisation in
Sec.~\ref{sec:polarisation_mond}. This indicates that the mass of
  the ghost $m_\text{BD}$ could be large.

\section{Conclusions}
\label{sec:conclusion}

In this work we followed the philosophy of combining the different
approaches used to tackle the cosmological constant and dark matter
problems. The origin of this dark sector constitutes one of the most
challenging puzzle of contemporary physics. Here we want essentially
to consider them on the same footing. The standard model of cosmology
$\Lambda$-CDM, despites many observational successes, fails to explain
the observed tiny value of the cosmological constant in the presence
of large quantum corrections using standard quantum field theory
techniques. This has initiated the development of IR modifications of
GR, like massive gravity and bigravity. On the other hand, the model
$\Lambda$-CDM does not account for many observations of dark matter at
galactic scales, being unable to explain without fine-tuning the tight
correlations between the dark and luminous matter in galaxy
halos. Rather, the dark matter phenomenology at galactic scales is in
good agreement with MOND~\cite{Milg1, Milg2, Milg3}.

In the present work we aimed at addressing these two motivations under
the same umbrella using a common framework, namely the one of
bigravity. An important clue in this respect is the fact that the MOND
acceleration scale is of the order of the cosmological constant,
$a_0\sim\sqrt{\Lambda}$. An additional hope was to be able to promote
the MOND formula into a decent relativistic theory in the context of
bigravity. For this purpose we followed tightly the same ingredients
as in the model proposed in~\cite{Blanchet:2015sra}, with two species
of dark matter particles coupled to the two metrics of bigravity
respectively, and linked \textit{via} an internal vector field.

This paper was dedicated to explore the theoretical and
phenomenological consistency of this model and verify that it is
capable to recover the MOND phenomenology on galactic scales. We first
worked out the covariant field equations with a special emphasis on
the contributions of the vector field and the dark matter
particles. Because of the interaction term between the dark matter
particles and the vector field, the stress-energy tensors are not
conserved separately, but rather a combination of them, giving rise to
a global conservation law. The divergence of the stress-energy tensor
with respect to the effective metric $g_{\text{eff}}$ to which is
coupled the vector field~\cite{deRham:2014naa, deRham:2014fha}, is
given by the interaction of the dark matter particles with the vector
field, which has important consequences for the polarisation mechanism
but also for our decoupling limit analysis.

As next we computed the linear field equations around a de Sitter
background where the mass term shall play the role of the cosmological
constant on large scales. On small scales where the post-Newtonian
limit applies, the de Sitter background can be approximated by a flat
Minkowski background. Considering a small perturbation of the
Minkowski metric and computing the Newtonian limit, we were able to
show that the polarisation mechanism works successfully and recovers
the MOND phenomenology on galactic scales. We find that this
polarisation mechanism is a natural consequence of bimetric gravity
since it is made possible by the constraint equation (linearized
Bianchi identity) linking together the two metrics of bigravity. In
addition it strongly relies on the internal vector field generated by
the coupling between the two species of DM particles living in the two
metrics, as it is alive due to an appropriate cancelation between the
gravitational force and the internal vector force.

We then studied the decoupling limit of the matter sector in
details. We found that precisely because of this coupling to the
vector field (whose importance is probably priceless for the
polarisation to work) a ghost is reintroduced in the dark matter
sector. For the particular solution we have for the dark matter the
ghost is higher order in the perturbative expansion. The exact
  mass of the ghost shall be studied in a future work.

\acknowledgments

We would like to thank Jose Beltran Jimenez, Claudia de Rham and
Gilles Esposito-Far\`ese for very useful discussions. L.H. wishes to
acknowledge the African Institute for Mathematical Sciences in
Muizenberg, South Africa, for hospitality and support at the latest
stage of this work.


	\bibliographystyle{JHEPmodplain}
	\bibliography{DDM_LB_LH}

\providecommand{\href}[2]{#2}\begingroup\raggedright\begin{thebibliography}{10}

\bibitem{OS95}
J.~Ostriker and P.~Steinhardt, {\it Cosmic concordance},  {\sl Nature} {\bf
  377} (1995) 600.

\bibitem{Padmanabhan}
T.~Padmanabhan, {\it Cosmological constant -- the weight of the vacuum},  {\sl
  Phys. Rept.} {\bf 380} (2003) 235, [\href{http://arxiv.org/abs/arXiv:0212290
  [hep-th]}{{\sf arXiv:0212290 [hep-th]}}].

\bibitem{Martin}
J.~Martin, {\it Everything you always wanted to know about the cosmological
  constant problem (but were afraid to ask)},  {\sl Comptes Rendus Physique}
  {\bf 13} (2012) 6, [\href{http://arxiv.org/abs/arXiv:1205.3365
  [astro-ph]}{{\sf arXiv:1205.3365 [astro-ph]}}].

\bibitem{deRham:2010ik}
C.~de~Rham and G.~Gabadadze, {\it {Generalization of the Fierz-Pauli action}},
  {\sl Phys. Rev.} {\bf D82} (2010) 044020,
  [\href{http://arxiv.org/abs/1007.0443}{{\sf arXiv:1007.0443}}].

\bibitem{deRham:2010kj}
C.~de~Rham, G.~Gabadadze, and A.~J. Tolley, {\it {Resummation of massive
  gravity}},  {\sl Phys. Rev. Lett.} {\bf 106} (2011) 231101,
  [\href{http://arxiv.org/abs/1011.1232}{{\sf arXiv:1011.1232}}].

\bibitem{deRham11}
C.~de~Rham, G.~Gabadadze, and A.~Tolley, {\it {Ghost free Massive Gravity in
  the St\"uckelberg language}},  {\sl Phys. Lett. B} {\bf 711} (2012) 190--195,
  [\href{http://arxiv.org/abs/arXiv:1107.3820 [hep-th]}{{\sf arXiv:1107.3820
  [hep-th]}}].

\bibitem{Hassan:2011hr}
S.~Hassan and R.~A. Rosen, {\it {Resolving the ghost problem in non-linear
  massive gravity}},  {\sl Phys. Rev. Lett.} {\bf 108} (2012) 041101,
  [\href{http://arxiv.org/abs/1106.3344}{{\sf arXiv:1106.3344}}].

\bibitem{Hassan:2011vm}
S.~Hassan and R.~A. Rosen, {\it {On Non-Linear Actions for Massive Gravity}},
  {\sl J. High Energy Phys.} {\bf 1107} (2011) 009,
  [\href{http://arxiv.org/abs/1103.6055}{{\sf arXiv:1103.6055}}],
  [\href{http://dx.doi.org/10.1007/JHEP07(2011)009}{{\sf
  doi:10.1007/JHEP07(2011)009}}].

\bibitem{Hassan:2011tf}
S.~Hassan, R.~A. Rosen, and A.~Schmidt-May, {\it {Ghost-free Massive Gravity
  with a General Reference Metric}},  {\sl JHEP} {\bf 1202} (2012) 026,
  [\href{http://arxiv.org/abs/1109.3230}{{\sf arXiv:1109.3230}}],
  [\href{http://dx.doi.org/10.1007/JHEP02(2012)026}{{\sf
  doi:10.1007/JHEP02(2012)026}}].

\bibitem{Hassan:2011ea}
S.~Hassan and R.~A. Rosen, {\it {Confirmation of the secondary constraint and
  absence of ghost in massive gravity and bimetric gravity}},  {\sl J. High
  Energy Phys.} {\bf 1204} (2012) 123,
  [\href{http://arxiv.org/abs/1111.2070}{{\sf arXiv:1111.2070}}].

\bibitem{Hassan:2012qv}
S.~Hassan, A.~Schmidt-May, and M.~von Strauss, {\it {Proof of Consistency of
  Nonlinear Massive Gravity in the St\'uckelberg Formulation}},  {\sl
  Phys.Lett.} {\bf B715} (2012) 335--339,
  [\href{http://arxiv.org/abs/1203.5283}{{\sf arXiv:1203.5283}}],
  [\href{http://dx.doi.org/10.1016/j.physletb.2012.07.018}{{\sf
  doi:10.1016/j.physletb.2012.07.018}}].

\bibitem{Boulware:1973my}
D.~Boulware and S.~Deser, {\it {Can gravitation have a finite range?}},  {\sl
  Phys. Rev.} {\bf D6} (1972) 3368--3382.

\bibitem{Hassan:2011zd}
S.~Hassan and R.~A. Rosen, {\it {Bimetric Gravity from Ghost-free Massive
  Gravity}},  {\sl J. High Energy Phys.} {\bf 1202} (2012) 126,
  [\href{http://arxiv.org/abs/1109.3515}{{\sf arXiv:1109.3515}}],
  [\href{http://dx.doi.org/10.1007/JHEP02(2012)126}{{\sf
  doi:10.1007/JHEP02(2012)126}}].

\bibitem{Yamashita:2014fga}
Y.~Yamashita, A.~De~Felice, and T.~Tanaka, {\it {Appearance of Boulware--Deser
  ghost in bigravity with doubly coupled matter}},  {\sl Int. J. Mod. Phys.}
  {\bf D23} (2014) 1443003, [\href{http://arxiv.org/abs/1408.0487}{{\sf
  arXiv:1408.0487}}], [\href{http://dx.doi.org/10.1142/S0218271814430032}{{\sf
  doi:10.1142/S0218271814430032}}].

\bibitem{deRham:2014naa}
C.~de~Rham, L.~Heisenberg, and R.~H. Ribeiro, {\it {On couplings to matter in
  massive (bi-)gravity}},  \href{http://arxiv.org/abs/1408.1678}{{\sf
  arXiv:1408.1678}}.

\bibitem{Hassan:2012wr}
S.~Hassan, A.~Schmidt-May, and M.~von Strauss, {\it {On Consistent Theories of
  Massive Spin-2 Fields Coupled to Gravity}},  {\sl JHEP} {\bf 1305} (2013)
  086, [\href{http://arxiv.org/abs/1208.1515}{{\sf arXiv:1208.1515}}],
  [\href{http://dx.doi.org/10.1007/JHEP05(2013)086}{{\sf
  doi:10.1007/JHEP05(2013)086}}].

\bibitem{deRham:2014fha}
C.~de~Rham, L.~Heisenberg, and R.~H. Ribeiro, {\it {Ghosts and Matter Couplings
  in Massive (bi-and multi-)Gravity}},  {\sl Phys. Rev.} {\bf D90} (2014)
  124042, [\href{http://arxiv.org/abs/1409.3834}{{\sf arXiv:1409.3834}}],
  [\href{http://dx.doi.org/10.1103/PhysRevD.90.124042}{{\sf
  doi:10.1103/PhysRevD.90.124042}}].

\bibitem{Noller:2014sta}
J.~Noller and S.~Melville, {\it {The coupling to matter in Massive, Bi- and
  Multi-Gravity}},  \href{http://arxiv.org/abs/1408.5131}{{\sf
  arXiv:1408.5131}}.

\bibitem{LH15}
L.~Heisenberg, {\it Quantum corrections in massive bigravity and new effective
  composite metrics},  {\sl Class. Quantum Grav.} {\bf 32} (2015) 105011,
  [\href{http://arxiv.org/abs/1410.4239}{{\sf arXiv:1410.4239}}].

\bibitem{deRham:2015cha}
C.~de~Rham and A.~J. Tolley, {\it {Vielbein to the Rescue?}},
  \href{http://arxiv.org/abs/1505.01450}{{\sf arXiv:1505.01450}}.

\bibitem{Huang:2015yga}
Q.-G. Huang, R.~H. Ribeiro, Y.-H. Xing, K.-C. Zhang, and S.-Y. Zhou, {\it {On
  the uniqueness of the non-minimal matter coupling in massive gravity and
  bigravity}},  \href{http://arxiv.org/abs/1505.02616}{{\sf arXiv:1505.02616}}.

\bibitem{Heisenberg:2015iqa}
L.~Heisenberg, {\it {More on effective composite metrics}},
  \href{http://arxiv.org/abs/1505.02966}{{\sf arXiv:1505.02966}}.

\bibitem{PhysRevD.84.124046}
G.~D'Amico, C.~de~Rham, S.~Dubovsky, G.~Gabadadze, D.~Pirtskhalava, and A.~J.
  Tolley, {\it Massive cosmologies},  {\sl Phys. Rev. D} {\bf 84} (Dec, 2011)
  124046, [\href{http://arxiv.org/abs/1108.5231}{{\sf 1108.5231}}].

\bibitem{deRham:2012ew}
C.~de~Rham, G.~Gabadadze, L.~Heisenberg, and D.~Pirtskhalava, {\it
  {Non-renormalization and naturalness in a class of scalar-tensor theories}},
  {\sl Phys. Rev.} {\bf D87} (2012) [\href{http://arxiv.org/abs/1212.4128}{{\sf
  arXiv:1212.4128}}].

\bibitem{Luty:2003vm}
M.~A. Luty, M.~Porrati, and R.~Rattazzi, {\it {Strong interactions and
  stability in the DGP model}},  {\sl J. High Energy Phys.} {\bf 0309} (2003)
  029, [\href{http://arxiv.org/abs/hep-th/0303116}{{\sf
  arXiv:hep-th/0303116}}].

\bibitem{Nicolis:2004qq}
A.~Nicolis and R.~Rattazzi, {\it {Classical and quantum consistency of the DGP
  model}},  {\sl J. High Energy Phys.} {\bf 0406} (2004) 059,
  [\href{http://arxiv.org/abs/hep-th/0404159}{{\sf arXiv:hep-th/0404159}}].

\bibitem{Heisenberg:2014raa}
L.~Heisenberg, {\it {Quantum corrections in Galileons from matter loops}},
  \href{http://arxiv.org/abs/1408.0267}{{\sf arXiv:1408.0267}}.

\bibitem{deRham:2013qqa}
C.~de~Rham, L.~Heisenberg, and R.~H. Ribeiro, {\it {Quantum Corrections in
  Massive Gravity}},  {\sl Phys. Rev.} {\bf D88} (2013) 084058,
  [\href{http://arxiv.org/abs/1307.7169}{{\sf arXiv:1307.7169}}],
  [\href{http://dx.doi.org/10.1103/PhysRevD.88.084058}{{\sf
  doi:10.1103/PhysRevD.88.084058}}].

\bibitem{Heisenberg:2014rka}
L.~Heisenberg, {\it {Quantum corrections in massive bigravity and new effective
  composite metrics}},  \href{http://arxiv.org/abs/1410.4239}{{\sf
  arXiv:1410.4239}}.

\bibitem{deRham:2010tw}
C.~de~Rham, G.~Gabadadze, L.~Heisenberg, and D.~Pirtskhalava, {\it {Cosmic
  acceleration and the helicity-0 graviton}},  {\sl Phys. Rev. D} {\bf 83}
  (2011) 103516, [\href{http://arxiv.org/abs/1010.1780}{{\sf
  arXiv:1010.1780}}].

\bibitem{Tasinato:2012ze}
G.~Tasinato, K.~Koyama, and G.~Niz, {\it {Vector instabilities and
  self-acceleration in the decoupling limit of massive gravity}},  {\sl Phys.
  Rev.} {\bf D87} (2013) 064029, [\href{http://arxiv.org/abs/1210.3627}{{\sf
  arXiv:1210.3627}}].

\bibitem{Gumrukcuoglu:2011ew}
A.~E. Gumrukcuoglu, C.~Lin, and S.~Mukohyama, {\it {Open FRW universes and
  self-acceleration from nonlinear massive gravity}},  {\sl J. Cosm.
  Astropart.} {\bf 1111} (2011) 030,
  [\href{http://arxiv.org/abs/1109.3845}{{\sf arXiv:1109.3845}}].

\bibitem{Gumrukcuoglu:2011zh}
A.~E. Gumrukcuoglu, C.~Lin, and S.~Mukohyama, {\it {Cosmological perturbations
  of self-accelerating universe in nonlinear massive gravity}},  {\sl J. Cosm.
  Astropart.} {\bf 1203} (2012) 006,
  [\href{http://arxiv.org/abs/1111.4107}{{\sf arXiv:1111.4107}}],
  [\href{http://dx.doi.org/10.1088/1475-7516/2012/03/006}{{\sf
  doi:10.1088/1475-7516/2012/03/006}}].

\bibitem{PhysRevLett.109.171101}
A.~De~Felice, A.~E. G\"umr\"uk\ifmmode \mbox{\c{c}}\else
  \c{c}\fi{}\"uo\ifmmode~\breve{g}\else \u{g}\fi{}lu, and S.~Mukohyama, {\it
  Massive gravity: Nonlinear instability of a homogeneous and isotropic
  universe},  {\sl Phys. Rev. Lett.} {\bf 109} (Oct, 2012) 171101,
  [\href{http://arxiv.org/abs/1206.2080}{{\sf arXiv:1206.2080}}],
  [\href{http://dx.doi.org/10.1103/PhysRevLett.109.171101}{{\sf
  doi:10.1103/PhysRevLett.109.171101}}].

\bibitem{Fasiello:2012rw}
M.~Fasiello and A.~J. Tolley, {\it {Cosmological perturbations in Massive
  Gravity and the Higuchi bound}},  {\sl J. Cosm. Astropart.} {\bf 1211} (2012)
  035, [\href{http://arxiv.org/abs/1206.3852}{{\sf arXiv:1206.3852}}],
  [\href{http://dx.doi.org/10.1088/1475-7516/2012/11/035}{{\sf
  doi:10.1088/1475-7516/2012/11/035}}].

\bibitem{Langlois:2012hk}
D.~Langlois and A.~Naruko, {\it {Cosmological solutions of massive gravity on
  de Sitter}},  {\sl Class.Quant.Grav.} {\bf 29} (2012) 202001,
  [\href{http://arxiv.org/abs/1206.6810}{{\sf arXiv:1206.6810}}],
  [\href{http://dx.doi.org/10.1088/0264-9381/29/20/202001}{{\sf
  doi:10.1088/0264-9381/29/20/202001}}].

\bibitem{Huang:2012pe}
Q.-G. Huang, Y.-S. Piao, and S.-Y. Zhou, {\it {Mass-Varying Massive Gravity}},
  {\sl Phys. Rev.} {\bf D86} (2012) 124014,
  [\href{http://arxiv.org/abs/1206.5678}{{\sf arXiv:1206.5678}}],
  [\href{http://dx.doi.org/10.1103/PhysRevD.86.124014}{{\sf
  doi:10.1103/PhysRevD.86.124014}}].

\bibitem{Mukohyama:2014rca}
S.~Mukohyama, {\it {A new quasidilaton theory of massive gravity}},  {\sl J.
  Cosm. Astropart.} {\bf 1412} (2014), no.~12 011,
  [\href{http://arxiv.org/abs/1410.1996}{{\sf arXiv:1410.1996}}],
  [\href{http://dx.doi.org/10.1088/1475-7516/2014/12/011}{{\sf
  doi:10.1088/1475-7516/2014/12/011}}].

\bibitem{Volkov:2011an}
M.~S. Volkov, {\it {Cosmological solutions with massive gravitons in the
  bigravity theory}},  {\sl JHEP} {\bf 1201} (2012) 035,
  [\href{http://arxiv.org/abs/1110.6153}{{\sf arXiv:1110.6153}}],
  [\href{http://dx.doi.org/10.1007/JHEP01(2012)035}{{\sf
  doi:10.1007/JHEP01(2012)035}}].

\bibitem{vonStrauss:2011mq}
M.~von Strauss, A.~Schmidt-May, J.~Enander, E.~Mortsell, and S.~Hassan, {\it
  {Cosmological Solutions in Bimetric Gravity and their Observational Tests}},
  {\sl JCAP} {\bf 1203} (2012) 042, [\href{http://arxiv.org/abs/1111.1655}{{\sf
  arXiv:1111.1655}}],
  [\href{http://dx.doi.org/10.1088/1475-7516/2012/03/042}{{\sf
  doi:10.1088/1475-7516/2012/03/042}}].

\bibitem{Comelli:2014bqa}
D.~Comelli, M.~Crisostomi, and L.~Pilo, {\it {FRW Cosmological Perturbations in
  Massive Bigravity}},  {\sl Phys. Rev.} {\bf D90} (2014) 084003,
  [\href{http://arxiv.org/abs/1403.5679}{{\sf arXiv:1403.5679}}],
  [\href{http://dx.doi.org/10.1103/PhysRevD.90.084003}{{\sf
  doi:10.1103/PhysRevD.90.084003}}].

\bibitem{Comelli:2012db}
D.~Comelli, M.~Crisostomi, and L.~Pilo, {\it {Perturbations in Massive Gravity
  Cosmology}},  {\sl J. High Energy Phys.} {\bf 1206} (2012) 085,
  [\href{http://arxiv.org/abs/1202.1986}{{\sf arXiv:1202.1986}}],
  [\href{http://dx.doi.org/10.1007/JHEP06(2012)085}{{\sf
  doi:10.1007/JHEP06(2012)085}}].

\bibitem{DeFelice:2013nba}
A.~De~Felice, T.~Nakamura, and T.~Tanaka, {\it {Possible existence of viable
  models of bi-gravity with detectable graviton oscillations by gravitational
  wave detectors}},  {\sl PTEP} {\bf 2014} (2014), no.~4 043E01,
  [\href{http://arxiv.org/abs/1304.3920}{{\sf arXiv:1304.3920}}],
  [\href{http://dx.doi.org/10.1093/ptep/ptu024}{{\sf
  doi:10.1093/ptep/ptu024}}].

\bibitem{DeFelice:2014nja}
A.~De~Felice, A.~E. G{\"u}mr{\"u}k{\c c}{\"u}o{\u g}lu, S.~Mukohyama,
  N.~Tanahashi, and T.~Tanaka, {\it {Viable cosmology in bimetric theory}},
  {\sl J. Cosm. Astropart.} {\bf 1406} (2014) 037,
  [\href{http://arxiv.org/abs/1404.0008}{{\sf arXiv:1404.0008}}],
  [\href{http://dx.doi.org/10.1088/1475-7516/2014/06/037}{{\sf
  doi:10.1088/1475-7516/2014/06/037}}].

\bibitem{Akrami:2015qga}
Y.~Akrami, S.~Hassan, F.~K{\"o}nnig, A.~Schmidt-May, and A.~R. Solomon, {\it
  {Bimetric gravity is cosmologically viable}},
  \href{http://arxiv.org/abs/1503.07521}{{\sf arXiv:1503.07521}},
  \href{http://dx.doi.org/10.1016/j.physletb.2015.06.062}{{\sf
  doi:10.1016/j.physletb.2015.06.062}}.

\bibitem{Enander:2015vja}
J.~Enander, Y.~Akrami, E.~M{\"o}rtsell, M.~Renneby, and A.~R. Solomon, {\it
  {Integrated Sachs-Wolfe effect in massive bigravity}},  {\sl Phys.Rev.} {\bf
  D91} (2015) 084046, [\href{http://arxiv.org/abs/1501.02140}{{\sf
  arXiv:1501.02140}}],
  [\href{http://dx.doi.org/10.1103/PhysRevD.91.084046}{{\sf
  doi:10.1103/PhysRevD.91.084046}}].

\bibitem{Enander:2015kda}
J.~Enander and E.~Mortsell, {\it {On stars, galaxies and black holes in massive
  bigravity}},  \href{http://arxiv.org/abs/1507.00912}{{\sf arXiv:1507.00912}}.

\bibitem{Cusin:2015pya}
G.~Cusin, R.~Durrer, P.~Guarato, and M.~Motta, {\it {Inflationary perturbations
  in bimetric gravity}},  \href{http://arxiv.org/abs/1505.01091}{{\sf
  arXiv:1505.01091}}.

\bibitem{Konnig:2015lfa}
F.~K{\"o}nnig, {\it {Higuchi Ghosts and Gradient Instabilities in Bimetric
  Gravity}},  {\sl Phys.Rev.} {\bf D91} (2015) 104019,
  [\href{http://arxiv.org/abs/1503.07436}{{\sf arXiv:1503.07436}}],
  [\href{http://dx.doi.org/10.1103/PhysRevD.91.104019}{{\sf
  doi:10.1103/PhysRevD.91.104019}}].

\bibitem{Fasiello:2015csa}
M.~Fasiello and R.~H. Ribeiro, {\it {Mild bounds on bigravity from primordial
  gravitational waves}},  \href{http://arxiv.org/abs/1505.00404}{{\sf
  arXiv:1505.00404}}.

\bibitem{Bosma}
A.~Bosma, {\it 21-cm line studies of spiral galaxies. ii. the distribution and
  kinematics of neutral hydrogen in spiral galaxies of various morphological
  types},  {\sl Astron. J.} {\bf 86} (1981) 1825.

\bibitem{Rubin}
V.~Rubin, J.~Ford, N.~Thonnard, and D.~Burstein, {\it Rotational properties of
  23 sb galaxies},  {\sl Astrophys. J.} {\bf 261} (1982) 439.

\bibitem{HuD02}
W.~Hu and S.~Dodelson, {\it Cosmic microwave background anisotropies},  {\sl
  Annual Rev. Astron. Astrophys.} {\bf 40} (2002) 171.

\bibitem{BHS05}
G.~Bertone, D.~Hooper, and J.~Silk, {\it Particle dark matter: Evidence,
  candidates and constraints},  {\sl Phys. Rept.} {\bf 405} (2005) 279,
  [\href{http://arxiv.org/abs/hep-ph/0404175}{{\sf hep-ph/0404175}}].

\bibitem{SandMcG02}
R.~Sanders and S.~McGaugh, {\it Modified newtonian dynamics as an alternative
  to dark matter},  {\sl Ann. Rev. Astron. Astrophys.} {\bf 40} (2002) 263,
  [\href{http://arxiv.org/abs/astro-ph/0204521}{{\sf astro-ph/0204521}}].

\bibitem{FamMcG12}
B.~Famaey and S.~McGaugh, {\it Modified {N}ewtonian dynamics ({MOND}):
  Observational phenomenology and relativistic extensions},  {\sl Living Rev.
  Rel.} {\bf 15} (2012) 10, [\href{http://arxiv.org/abs/arXiv:1112.3960
  [astro-ph.CO]}{{\sf arXiv:1112.3960 [astro-ph.CO]}}].

\bibitem{TF77}
R.~Tully and J.~Fisher, {\it A new method of determining distances to
  galaxies},  {\sl Astron. Astrophys.} {\bf 54} (1977) 661.

\bibitem{McG00}
S.~McGaugh, J.~Schombert, G.~Bothun, and W.~de~Blok, {\it The baryonic
  tully-fisher relation},  {\sl Astrophys. J.} {\bf 533} (2000), no.~2 L99.

\bibitem{McG11}
S.~S. McGaugh, {\it Novel test of modified newtonian dynamics with gas rich
  galaxies},  {\sl Phys. Rev. Lett.} {\bf 106} (2011) 121303.

\bibitem{Sand10}
R.~Sanders, {\it The universal faber-jackson relation},  {\sl Mon. Not. Roy.
  Astron. Soc.} {\bf 407} (2010) 1128--1134,
  [\href{http://arxiv.org/abs/arXiv:1002.2765 [astro-ph.CO]}{{\sf
  arXiv:1002.2765 [astro-ph.CO]}}].

\bibitem{Milg1}
M.~Milgrom, {\it A modification of the {N}ewtonian dynamics as a possible
  alternative to the hidden mass hypothesis},  {\sl Astrophys. J.} {\bf 270}
  (1983) 365.

\bibitem{Milg2}
M.~Milgrom, {\it A modification of the {N}ewtonian dynamics: Implications for
  galaxies},  {\sl Astrophys. J.} {\bf 270} (1983) 371.

\bibitem{Milg3}
M.~Milgrom, {\it A modification of the {N}ewtonian dynamics: Implications for
  galaxy systems},  {\sl Astrophys. J.} {\bf 270} (1983) 384.

\bibitem{BekM84}
J.~Bekenstein and M.~Milgrom, {\it Does the missing mass problem signal the
  breakdown of newtonian gravity?},  {\sl Astrophys. J.} {\bf 286} (1984) 7.

\bibitem{GD92}
D.~Gerbal, F.~Durret, M.~Lachi\`eze-Rey, and G.~Lima-Neto, {\it Analysis of
  x-ray galaxy clusters in the framework of modified newtonian dynamics},  {\sl
  Astron. Astrophys.} {\bf 262} (1992) 395.

\bibitem{PSilk05}
E.~Pointecouteau and J.~Silk, {\it New constraints on modified newtonian
  dynamics from galaxy clusters},  {\sl Mon. Not. R. Astron. Soc.} {\bf 364}
  (2005) 654--658, [\href{http://arxiv.org/abs/astro-ph/0505017}{{\sf
  astro-ph/0505017}}].

\bibitem{Clowe06}
D.~Clowe, M.~Bradac, A.~H. Gonzalez, M.~Markevitch, S.~W. Randall, C.~Jones,
  and D.~Zaritsky, {\it A direct empirical proof of the existence of dark
  matter},  {\sl Astrophys. J.} {\bf 648} (2006) L109,
  [\href{http://arxiv.org/abs/astro-ph/0608407}{{\sf astro-ph/0608407}}].

\bibitem{Ang08}
G.~W. Angus, B.~Famaey, and D.~A. Buote, {\it {X}-ray group and cluster mass
  profiles in {MOND}: Unexplained mass on the group scale},  {\sl Mon. Not.
  Roy. Astron. Soc.} {\bf 387} (2008) 1470,
  [\href{http://arxiv.org/abs/arXiv:0709.0108 [astro-ph]}{{\sf arXiv:0709.0108
  [astro-ph]}}].

\bibitem{Ang09}
G.~W. Angus, {\it Is an 11 ev sterile neutrino consistent with clusters, the
  cosmic microwave background and modified {N}ewtonian dynamics?},  {\sl Mon.
  Not. Roy. Astron. Soc.} {\bf 394} (2009) 527,
  [\href{http://arxiv.org/abs/arXiv:0805.4014 [astro-ph]}{{\sf arXiv:0805.4014
  [astro-ph]}}].

\bibitem{Milg09}
M.~Milgrom, {\it Mond effects in the inner solar system},  {\sl Mon. Not. Roy.
  Astron. Soc.} {\bf 399} (2009) 474.

\bibitem{BN11}
L.~Blanchet and J.~Novak, {\it External field effect of modified newtonian
  dynamics in the solar system},  {\sl Mon. Not. Roy. Astron. Soc.} {\bf 412}
  (2011) 2530.

\bibitem{Sand97}
R.~Sanders, {\it A stratified framework for scalar-tensor theories of modified
  dynamics},  {\sl Astrophys. J.} {\bf 480} (1997) 492,
  [\href{http://arxiv.org/abs/astro-ph/9612099}{{\sf astro-ph/9612099}}].

\bibitem{Bek04}
J.~Bekenstein, {\it Relativistic gravitation theory for the modified
  {N}ewtonian dynamics paradigm},  {\sl Phys. Rev. D} {\bf 70} (2004) 083509,
  [\href{http://arxiv.org/abs/astro-ph/0403694}{{\sf astro-ph/0403694}}].

\bibitem{Sand05}
R.~Sanders, {\it A tensor-vector-scalar framework for modified dynamics and
  cosmic dark matter},  {\sl Mon. Not. Roy. Astron. Soc.} {\bf 363} (2005) 459,
  [\href{http://arxiv.org/abs/astro-ph/0502222}{{\sf astro-ph/0502222}}].

\bibitem{ZFS07}
T.~G. Zlosnik, P.~G. Ferreira, and G.~D. Starkman, {\it Modifying gravity with
  the \ae ther: An alternative to dark matter},  {\sl Phys. Rev. D} {\bf 75}
  (2007) 044017, [\href{http://arxiv.org/abs/arXiv:astro-ph/0607411}{{\sf
  arXiv:astro-ph/0607411}}].

\bibitem{Halle08}
A.~Halle, H.~S. Zhao, and B.~Li, {\it A nonuniform dark energy fluid:
  Perturbation equations},  {\sl Astrophys. J. Suppl.} {\bf 177} (2008) 1,
  [\href{http://arxiv.org/abs/arXiv:0711.0958 [astro-ph]}{{\sf arXiv:0711.0958
  [astro-ph]}}].

\bibitem{bimond1}
M.~Milgrom, {\it Bimetric mond gravity},  {\sl Phys. Rev. D} {\bf 80} (2009)
  123536.

\bibitem{BDgef11}
E.~Babichev, C.~Deffayet, and G.~Esposito-Far{\`e}se, {\it Improving
  relativistic mond with galileon k-mouflage},  {\sl Phys. Rev. D} {\bf 84}
  (2011) 061502(R), [\href{http://arxiv.org/abs/arXiv:1106.2538 [gr-qc]}{{\sf
  arXiv:1106.2538 [gr-qc]}}].

\bibitem{DEW11}
C.~Deffayet, G.~Esposito-Far\`ese, and R.~Woodard, {\it Nonlocal metric
  formulations of mond with sufficient lensing},  {\sl Phys. Rev. D} {\bf 84}
  (2011) 124054, [\href{http://arxiv.org/abs/arXiv:1106.4984 [gr-qc]}{{\sf
  arXiv:1106.4984 [gr-qc]}}].

\bibitem{BM11}
L.~Blanchet and S.~Marsat, {\it Modified gravity approach based on a preferred
  time foliation},  {\sl Phys. Rev. D} {\bf 84} (2011) 044056,
  [\href{http://arxiv.org/abs/arXiv:1107.5264 [gr-qc]}{{\sf arXiv:1107.5264
  [gr-qc]}}].

\bibitem{Arraut2014}
Y.~Arraut, {\it Can a nonlocal model of gravity reproduce dark matter effects
  in agreement with mond?},  {\sl Int. J. of Modern Phys. D} {\bf 23} (2014)
  1450008, [\href{http://arxiv.org/abs/1310.0675}{{\sf arXiv:1310.0675}}].

\bibitem{B07mond}
L.~Blanchet, {\it Gravitational polarization and the phenomenology of {MOND}},
  {\sl Class. Quant. Grav.} {\bf 24} (2007) 3529,
  [\href{http://arxiv.org/abs/astro-ph/0605637}{{\sf astro-ph/0605637}}].

\bibitem{BL08}
L.~Blanchet and A.~Le~Tiec, {\it Model of dark matter and dark energy based on
  gravitational polarization},  {\sl Phys. Rev. D} {\bf 78} (2008) 024031,
  [\href{http://arxiv.org/abs/astro-ph/0804.3518}{{\sf astro-ph/0804.3518}}].

\bibitem{BL09}
L.~Blanchet and A.~Le~Tiec, {\it Dipolar dark matter and dark energy},  {\sl
  Phys. Rev. D} {\bf 80} (2009) 023524,
  [\href{http://arxiv.org/abs/arXiv:0901.3114 [astro-ph]}{{\sf arXiv:0901.3114
  [astro-ph]}}].

\bibitem{BLLM13}
L.~Blanchet, D.~Langlois, A.~Le~Tiec, and S.~Marsat, {\it Non-gaussianity in
  the cosmic microwave background induced by dipolar dark matter},  {\sl J.
  Cosm. Astropart.} {\bf 22} (2013) 1302,
  [\href{http://arxiv.org/abs/arXiv:1210.4106 [astro-ph]}{{\sf arXiv:1210.4106
  [astro-ph]}}].

\bibitem{BBwag}
L.~Blanchet and L.~Bernard, {\it Phenomenology of mond and gravitational
  polarization},  \href{http://arxiv.org/abs/arXiv:1403.5963 [gr-qc]}{{\sf
  arXiv:1403.5963 [gr-qc]}}. In the proceedings of the second Workshop on
  Antimatter and Gravity (WAG 2013).

\bibitem{BB14}
L.~Bernard and L.~Blanchet, {\it Phenomenology of dark matter via a bimetric
  extension of general relativity},  \href{http://arxiv.org/abs/arXiv:1410.7708
  [astro-ph]}{{\sf arXiv:1410.7708 [astro-ph]}}. To appear in Phys. Rev. D.

\bibitem{Blanchet:2015sra}
L.~Blanchet and L.~Heisenberg, {\it Dark matter via massive (bi-)gravity},
  \href{http://arxiv.org/abs/1504.00870}{{\sf arXiv:1504.00870}}. To appear in
  Phys. Rev. D.

\bibitem{Baccetti:2012re}
V.~Baccetti, P.~Martin-Moruno, and M.~Visser, {\it {Null Energy Condition
  violations in bimetric gravity}},  {\sl JHEP} {\bf 1208} (2012) 148,
  [\href{http://arxiv.org/abs/1206.3814}{{\sf arXiv:1206.3814}}],
  [\href{http://dx.doi.org/10.1007/JHEP08(2012)148}{{\sf
  doi:10.1007/JHEP08(2012)148}}].

\bibitem{Schmidt-May:2014xla}
A.~Schmidt-May, {\it {Mass eigenstates in bimetric theory with ghost-free
  matter coupling}},  \href{http://arxiv.org/abs/1409.3146}{{\sf
  arXiv:1409.3146}}.

\end{thebibliography}\endgroup

\end{document}